\documentclass[aps,prb,twocolumn,preprintnumbers,amsmath,amssymb,superscriptaddress]{revtex4}%

\usepackage{graphicx}%
\usepackage{dcolumn}
\usepackage{amsmath}
\usepackage{color}
\usepackage{multirow}
\usepackage{lmodern,bm}

\begin{document}

\title{Magnetism and its coexistence with superconductivity in CaK(Fe$_{0.949}$Ni$_{0.051}$)$_4$As$_4$: muon spin rotation/relaxation studies}
\author{Rustem~Khasanov}
 \email{rustem.khasanov@psi.ch}
 \affiliation{Laboratory for Muon Spin Spectroscopy, Paul Scherrer Institut, CH-5232 Villigen PSI, Switzerland}
\author{Gediminas Simutis}
 \affiliation{Laboratory for Muon Spin Spectroscopy, Paul Scherrer Institut, CH-5232 Villigen PSI, Switzerland}
 \affiliation{Laboratoire de Physique des Solides, Paris-Saclay University and CNRS}
\author{Yurii G. Pashkevich}
 \affiliation{O. O. Galkin Donetsk Institute for Physics and Engineering NAS of Ukraine, 03680 Kyiv-Kharkiv, Ukraine}
\author{Tatyana Shevtsova}
 \affiliation{O. O. Galkin Donetsk Institute for Physics and Engineering NAS of Ukraine, 03680 Kyiv-Kharkiv, Ukraine}
\author{William R. Meier}
 \affiliation{Division of Materials Science and Engineering, Ames Laboratory, Ames, Iowa 50011, USA}
 \affiliation{Department of Physics and Astronomy, Iowa State University, Ames, Iowa 50011, USA}
\author{Mingyu Xu}
 \affiliation{Division of Materials Science and Engineering, Ames Laboratory, Ames, Iowa 50011, USA}
 \affiliation{Department of Physics and Astronomy, Iowa State University, Ames, Iowa 50011, USA}
 
\author{Sergey L. Bud'ko}
 \affiliation{Division of Materials Science and Engineering, Ames Laboratory, Ames, Iowa 50011, USA}
 \affiliation{Department of Physics and Astronomy, Iowa State University, Ames, Iowa 50011, USA}
\author{Vladimir G. Kogan}
 \affiliation{Division of Materials Science and Engineering, Ames Laboratory, Ames, Iowa 50011, USA}
\author{Paul C. Canfield}
 \affiliation{Division of Materials Science and Engineering, Ames Laboratory, Ames, Iowa 50011, USA}
 \affiliation{Department of Physics and Astronomy, Iowa State University, Ames, Iowa 50011, USA}

\begin{abstract}
The magnetic response of CaK(Fe$_{0.949}$Ni$_{0.051}$)$_4$As$_4$ was investigated by means of the muon-spin rotation/relaxation. The long-range commensurate magnetic order sets in below the N\'{e}el temperature $T_{\rm N}= 50.0(5)$~K. The density-functional theory  calculations have identified three possible muon stopping sites. The experimental data were found to be consistent with only one type of magnetic structure, namely, the long-range magnetic spin-vortex-crystal order with the hedgehog motif within the $ab-$plane and the antiferromagnetic stacking along the $c-$direction. The value of the ordered magnetic moment at $T\approx3$~K was estimated to be $m_{\rm Fe}=0.38(11)$~$\mu_{\rm B}$ ($\mu_{\rm B}$ is the Bohr magneton). A microscopic coexistence of magnetic and superconducting phases accompanied by a reduction of the magnetic order parameter below the superconducting transition temperature $T_{\rm c}\simeq 9$~K is observed. Comparison with 11, 122, and 1144 families of Fe-based pnictides points to existence of correlation between the reduction of the magnetic order parameter at $T\rightarrow 0$ and the ratio of the transition temperatures $T_{\rm c}/T_{\rm N}$. Such correlations were found to be described by Machida's model for coexistence of itinerant spin-density wave magnetism and superconductivity [Machida, J. Phys. Soc. Jpn. {\bf 50}, 2195 (1981) and Bud'ko {\it et al.}, Phys. Rev. B {\bf 98}, 144520 (2018)].
\end{abstract}

%\pacs{74.70.Xa, 74.25.Bt, 74.45.+c, 76.75.+i}
\maketitle

\section{introduction}

\begin{figure}[htb]
\centering
\includegraphics[width=0.7\linewidth]{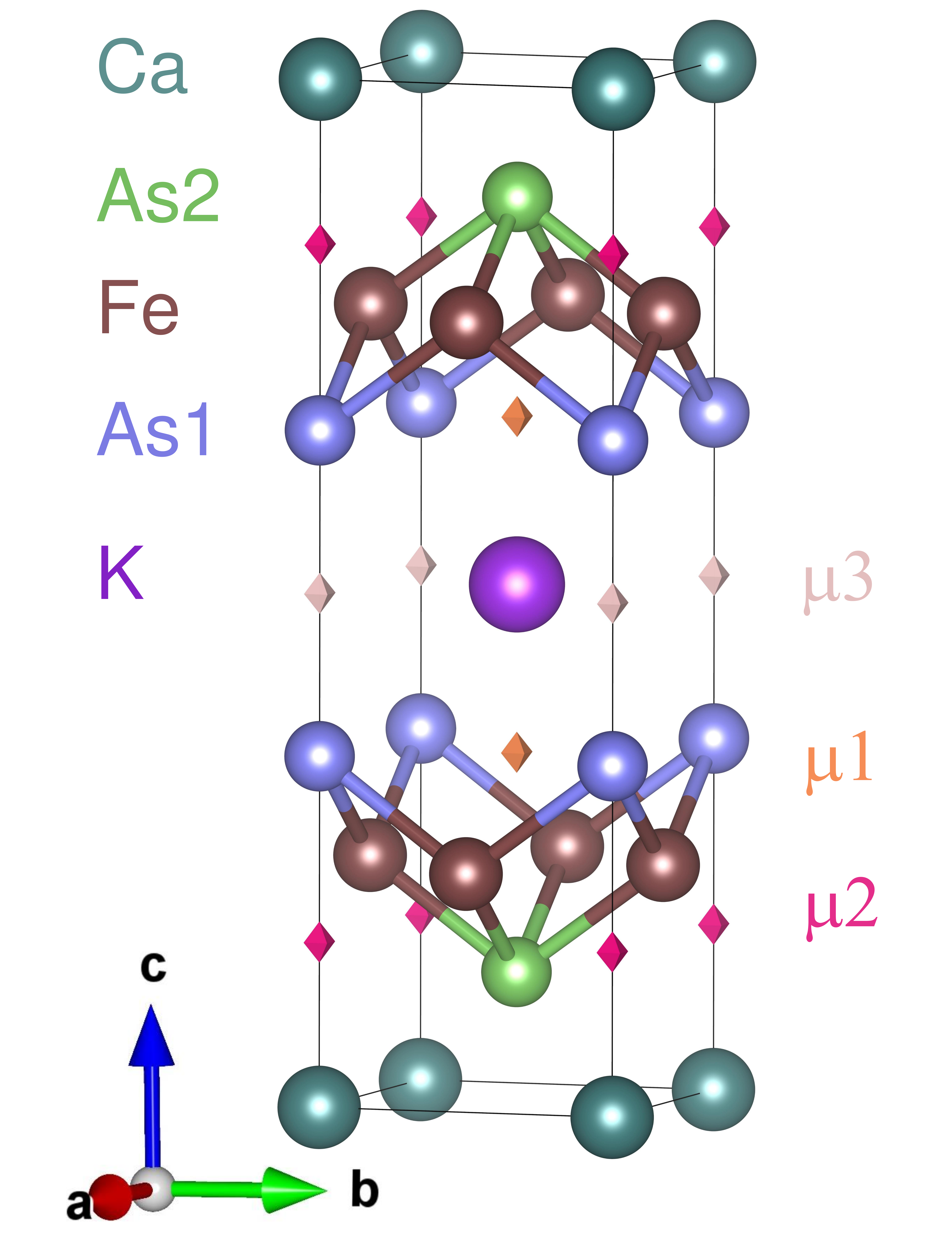}
% \vspace{-1.0cm}
%
\caption{ The crystal structure of CaKFe$_4$As$_4$ within the $P4/mmm$ group representation. Three muon stopping sites (denoted by diamonds) are: $\mu1$, the Wyckoff position is (0.5,0.5,0.301), the local symmetry is $2h$; $\mu2$, the Wyckoff position is (0,0,0.186), the local symmetry is $2g$; and $\mu 3$, the Wyckoff position is (0,0,0.5), the local symmetry is $1b$. The crystal structure is visualized by using VESTA package, Ref.~\onlinecite{Vesta}.}
 \label{fig:crystal-structure}
\end{figure}

Since their discovery, iron based superconductors (Fe-SC's) have attracted much interest. The materials belonging to various classes of Fe-SC's were found to be characterized  by unconventional superconducting properties, as well as by a strong interplay of superconductivity with various electronic ground states, such as {\it e.g.} nematic phase and spin-density wave magnetism.\cite{Kamihara_JAPS_2008, Hsu_PNAS_2008, Stewart_RMP_2011, Chen_NSR_2014, Paglione_NatPh_2010, Khasanov_La1111_PRB_2011, Khasanov_FeSe_PRB_2018} All Fe-SC's have a layered structure and share a common Fe$_2{\it An}_2$ ({\it An} = P, As, Se, Te) layers (see also Fig.~~\ref{fig:crystal-structure}), analogous to the CuO$_2$ sheets in high-temperature cuprates.\cite{Bednorz_ZPB_1986}

Recently a new, 1144 Fe-SC family ({\it AeA}Fe$_4$As$_4$, {\it Ae} = Ca, Sr, Eu and $A$ = K, Rb, Cs), with the transition temperature  $T_{\rm c}$ reaching $\simeq 36$~K, was synthesized.\cite{Iyo_JAPS_2016, Meier_PRB_2016, Meier_PRM_2017, Meier_PhD-Thesis_2018} The crystallographic structure of ${\it AeA}$Fe$_4$As$_4$ (see {\it e.g.} Fig.~\ref{fig:crystal-structure} for CaKFe$_4$As$_4$ representative of 1144 family) is different compared to intensively studied materials belonging to 122 family of Fe-SC's. The {\it Ae} and {\it A} sites form alternating planes along the crystallographic $c-$axis and are separated by Fe$_2$As$_2$ layers. There are two distinct As sites, As1 and As2, neighboring K and Ca, respectively, rather than one As site found in CaFe$_2$As$_2$ and KFe$_2$As$_2$ (see Fig.~\ref{fig:crystal-structure}).
Partial substitution of Fe by Co or Ni in CaKFe$_4$As$_4$ (electron doping) shift the ground state from superconducting to antiferromagnetically (AFM) ordered.\cite{Meier_PRM_2017, Meier_NPJ_2018, Meier_PhD-Thesis_2018}
The resistivity and  the specific heat measurements,\cite{Meier_PRB_2016, Meier_PRM_2017, Meier_PhD-Thesis_2018} as well as the M\"{o}ssbauer,\cite{Meier_NPJ_2018, Budko_PRB_2018} the nuclear magnetic resonance (NMR),\cite{Meier_NPJ_2018} and the neutron scattering studies\cite{Kreyssig_PRB_2018} reveal the appearance of a magnetic order.  The magnetism was further identified as the hedgehog spin-vortex-crystal (SVC) order, which is characterized by noncollinear Fe moments featuring an alternating all-in and all-out motif around the As1 sites within the $ab-$planes and antiferromagnetic coupling along the $c-$direction.\cite{Meier_NPJ_2018, Kreyssig_PRB_2018}
In addition, the interplay of magnetism and superconductivity in CaK(Fe$_{1-x}$Ni$_x)_4$As$_4$ was clearly detected in NMR, M\"{o}ssbauer spectroscopy and neutron scattering experiments.\cite{Meier_PRB_2016, Meier_PRM_2017, Meier_NPJ_2018, Meier_PhD-Thesis_2018, Kreyssig_PRB_2018, Budko_PRB_2018, Ding_PRB_2017} As both states compete for the same electrons at the Fermi surface, the magnetic order parameter was found to be strongly reduced below $T_{\rm c}$.

This paper reports the results of the muon-spin rotation/relaxation ($\mu$SR) studies of CaK(Fe$_{0.949}$Ni$_{0.051}$)$_4$As$_4$ single crystal sample. The density-functional theory  calculations have identified three possible muon stoping sites. The experimental data were found to be consistent with only one type of the magnetic structure, namely, the long-range magnetic spin-vortex-crystal order with the hedgehog motif within the $ab-$plane and the AFM stacking along the $c-$direction. The value of the ordered magnetic moment at $T\approx 3$~K was estimated to be $m_{\rm Fe}=0.38(11)$~$\mu_{\rm B}$ ($\mu_{\rm B}$ is the Bohr magneton). All these results stay in agreement with those published previously in Refs.~\onlinecite{Meier_NPJ_2018, Meier_PhD-Thesis_2018, Kreyssig_PRB_2018, Budko_PRB_2018}.
The temperature evolution of the magnetic order parameter $m_{\rm Fe}(T)$ was analyzed by using the approach of Machida,\cite{Machida_JPSJ_1981, Budko_PRB_2018} accounting for coexistence of a spin-density wave magnetism and superconductivity.
The theory results of Refs.~\onlinecite{Machida_JPSJ_1981, Budko_PRB_2018} were further extended for studying the superconductivity induced suppression of the magnetic order parameter in the limit of $T\rightarrow 0$.

The paper is organized as follows. The experimental details, including the sample preparation procedure, the description of $\mu$SR setup and the details of the muon-site calculations are given in Sec.~\ref{seq:experimental-details}. The results of the weak transverse-field and zero-field $\mu$SR experiments are presented in Sec.~\ref{sec:results}. Section \ref{seq:discussion} discusses the experimental data. Conclusions follow in Sec.~\ref{seq:conclusions}. Appendixes~\ref{sec:Symmetry_Analysis} and \ref{sec:Machida-Approach} describe the results of calculations of the dipolar fields at the muon stopping sites and extensions of Machida's theory at the limit of $T\rightarrow 0$, respectively.

\section{Experimental details \label{seq:experimental-details}}

\subsection{Sample preparation and characterization}

CaK(Fe$_{0.949}$Ni$_{0.051}$)$_4$As$_4$ single crystals were grown from a high-temperature Fe-As rich melt and extensively characterized by thermodynamic and transport measurements.\cite{Meier_PRB_2016, Meier_PRM_2017, Meier_PhD-Thesis_2018, Meier_NPJ_2018, Budko_PRB_2018, Kreyssig_PRB_2018} The selected  crystal with dimensions of $\simeq 4.0 \; \times \; 4.0 \; \times \; 0.1$~mm$^3$ was used. The magnetic ordering temperature $T_{\rm N}\simeq 50.6(5)$ and the superconducting transition temperature $T_{\rm c}\simeq 9.0(8)$~K, for CaK(Fe$_{0.949}$Ni$_{0.051}$)$_4$As$_4$ single crystals from the same grown batch, were inferred from temperature-dependent electrical-resistance, heat-capacity, and magnetization studies of Kreyssig {\it et al}.\cite{Kreyssig_PRB_2018}

\subsection{Muon-spin rotation/relaxation experiments } \label{seq:muSR}

\begin{figure}[htb]
%\centering
\includegraphics[width=1.0\linewidth]{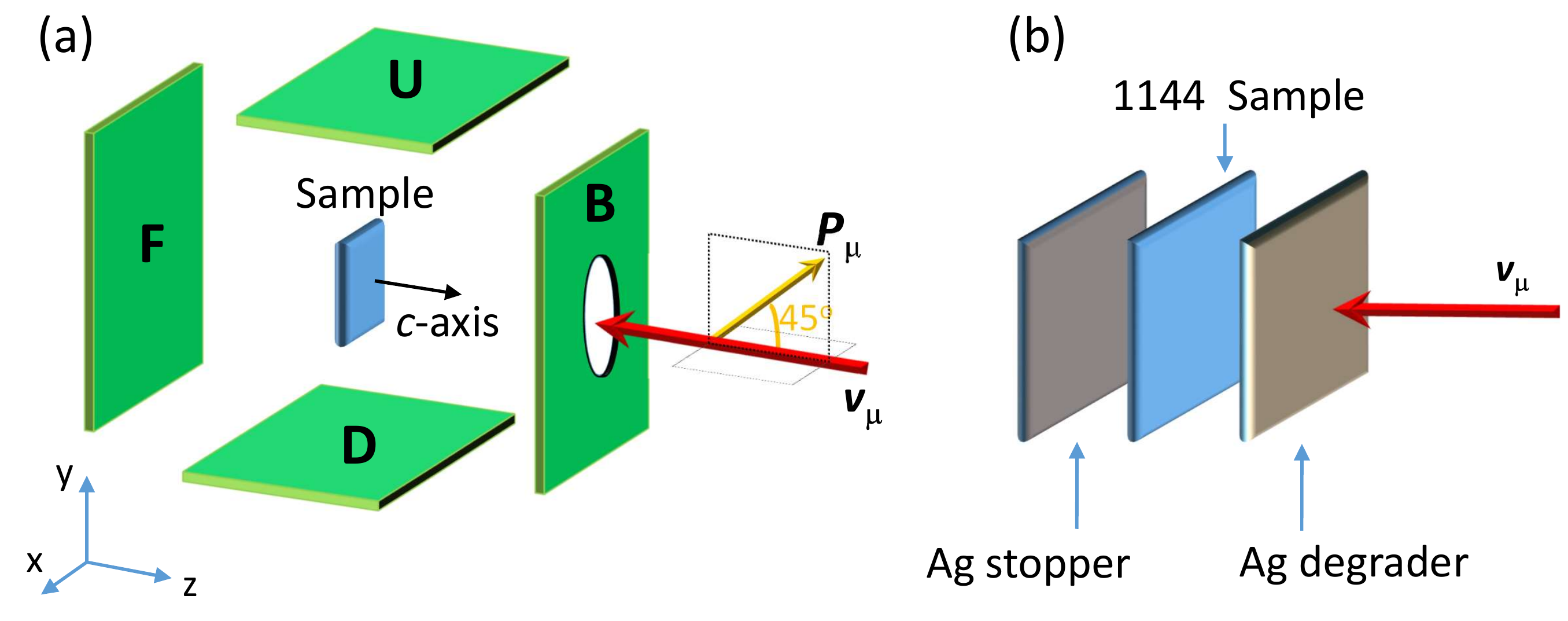}
% \vspace{-1.0cm}
%
\caption{(a) The schematic representation of an arrangement of positron counters at the GPS (General Purpose Surface) spectrometer at the Paul Scherrer institute, Switzerland.\cite{Amato_RSI_2017} The letters F,B, U, and D denote the Forward, Backward, Up, and Down detectors, respectively.  The single-crystalline sample (blue rectangle), has its $c-$axis aligned along the incoming muon beam (red arrow). The initial muon-spin polarization $P_\mu$  was rotated by 45$^{\rm o}$ within the vertical ($y-z$) plane. (b) CaK(Fe$_{0.949}$Ni$_{0.051}$)$_4$As$_4$ sample sandwiched between Ag (degrader and stopper) sheets. }
 \label{fig:muSR_setup}
\end{figure}

The muon-spin rotation/relaxation ($\mu$SR) experiments were carried out at the $\pi$M3 beam-line using the GPS (General Purpose Surface) spectrometer at the Paul Scherrer Institute, Switzerland.\cite{Amato_RSI_2017} The zero-field (ZF) and the weak transverse-field (wTF) $\mu$SR measurements were performed at temperatures ranging from $\simeq$1.5 to 100~K.
The 100\% spin-polarized muons with the momentum of $\simeq28.6$ MeV/c were implanted into the crystal along the $c-$axis (see Fig.~\ref{fig:muSR_setup}). Muons thermalize rapidly without a significant loss of their initial spin-polarization and stop in the matter at the depth of about 0.15 g/cm$^2$.  For CaK(Fe$_{0.949}$Ni$_{0.051}$)$_4$As$_4$ with the density of $\simeq5.2$~g/cm$^3$ this corresponds to a depth of $\simeq0.3$~mm. In order to measure the sample with a thickness of $\simeq0.1$~mm, as CaK(Fe$_{0.949}$Ni$_{0.051}$)$_4$As$_4$ single crystal studied here, a special sample holder described in Ref.~\onlinecite{Khasanov_FeSe_int_PRB_2016} was used. The sample was sandwiched between silver sheets, with the first one playing a role of a “degrader” by decelerating the muons in the incoming muon beam and the second one as a "stopper" by stopping the muons which were still able to penetrate through the sample [see Fig.~\ref{fig:muSR_setup}~(b)]. The detailed description of $\mu$SR technique can be found {\it e.g.} in Refs.~\onlinecite{Schenck_book_1985, Lee_book_1999, Brewer_book_1994, Yaouanc_book_2011}.

The $\mu$SR experiments were performed in the so-called spin-rotated mode, {\it i.e.} when the initial muon-spin polarization ($P_{\mu}$) is turned by a certain angle relative to the muon beam momentum [$v_\mu$, see Fig.~\ref{fig:muSR_setup}~(a)]. Such geometry is particularly suitable to perform experiments on single crystalline samples, since it allows to probe independently the time evolution of the 'parallel' and 'perpendicular' components of the muon-spin polarization [$P(t)$] by accessing the response of the Backward/Forward (B/F) and Up/Down (U/D) positron counters. In a case of the single crystalline CaK(Fe$_{0.949}$Ni$_{0.051}$)$_4$As$_4$ sample with the $c-$axis aligned along the muon momentum (see Fig.~\ref{fig:muSR_setup}) the Up/Down and Backward/Forward counters access the $P^{\perp c}(t)$ and $P^{\parallel c}(t)$ components of the muon-spin polarization, respectively.

The experimental data were analyzed by using the MUSRFIT package.\cite{MUSRFIT} The typical counting statistics were $\simeq 2\times 10^7$ and $3\times 10^6 $ positron events for ZF and wTF experiments, respectively.

\subsection{The muon stopping positions} \label{seq:stopping-sites}

The muon stopping sites can be identified as a local interstitial minima of the valence electron electrostatic potential which can be further restored from the electron density distribution. In order to find them, ab-initio calculations within the framework of density functional theory were performed. As the initial structural data, the $P4/mmm$ space group symmetry of CaKFe$_4$As$_4$ with one formula unit ($Z=1$) was considered (see Fig.~\ref{fig:crystal-structure}). The Ca ions reside at $1a$ Wyckoff position (0,0,0),  K at the $1d$ position (0.5,0.5,0.5), Fe at the $4i$ position (0,0.5,0.76820), As1 at the $2g$ position (0,0,0.34150), and As2 at the $2h$ position (0.5,0.5,0.12310).  The lattice constants were taken to be $a=3.8659$~$\AA$ and $c=12.8840$~$\AA$.\cite{Meier_PRM_2017,Meier_PhD-Thesis_2018} The all-electron full-potential linearized augmented plane wave method (Elk code),\cite{ELK-code} with the local spin-density approximation,\cite{Perdew_PRB_1992} for the exchange correlation potential and with the revised generalized gradient approximation of Perdew-Burke-Ernzerhof, \cite{Perdew_PRL_2008} was applied. The calculations were performed on a $13\times13\times4$ grid which corresponds to 84 points in the irreducible Brillouin zone.

Three types of possible muon positions were detected (see Fig.~\ref{fig:crystal-structure}).
The first one ($\mu 1$) stays in between K-As2 ions and has coordinates (0.5,0.5,0.301). The local symmetry of this position is $2h$. Two other positions are located on the line along the $c-$direction connecting the nearest Ca--As1-- As1--Ca ions. The local symmetries and positions are: $\mu2$ -- $2g$ (0,0,0.186) and $\mu3$ -- $1b$ (0,0,0.5). The positions $\mu1$ and $\mu2$ have the same local symmetry ($2g$ and $2h$) as they were found in Ba$_{1-x}A_x$Fe$_2$As$_2$.\cite{Mallett_EPL_2015, Sheveleva_Arxiv_2020}
The occupancy of the muon sites decreases from the site $\mu1$ to the site $\mu3$ ($\mu1 \rightarrow \mu2 \rightarrow \mu3$), as it follows from the ratio of corresponding electrostatic potentials: $(-\varphi_{\mu 1}) /(- \varphi_{\mu 2} )/(-\varphi_{\mu 3})\simeq 1.00/0.96/0.93$.
Two points need to be mentioned:
 (i) Few muon stopping positions with a much higher electrostatic potentials were assumed to be unoccupied by muons and, therefore, not considered.
 (ii) It is assumed that the muon sites calculated for CaKFe$_4$As$_4$ do not strongly change under the low level of Fe to Ni replacement.

\section{Results \label{sec:results}}

\subsection{Weak transverse-field $\mu$SR experiments} \label{seq:wTF}

$\mu$SR experiments under weak transverse-field (wTF) applied perpendicular to the muon-spin polarization are a straightforward method to determine the onset of the magnetic transition and the magnetic volume fraction. In this case the contribution to the asymmetry from muons experiencing a vanishing internal spontaneous magnetisation can be accurately determined. Muons stopping in a non-magnetic environment produce long lived oscillations, which reflect the coherent muon precession around the external field $B_{ex}$. Muons stopping in magnetically ordered parts of the sample give rise to a more complex, distinguishable signal, reflecting the vector combination of internal and external fields.

\begin{figure}[htb]
\centering
\includegraphics[width=0.8\linewidth]{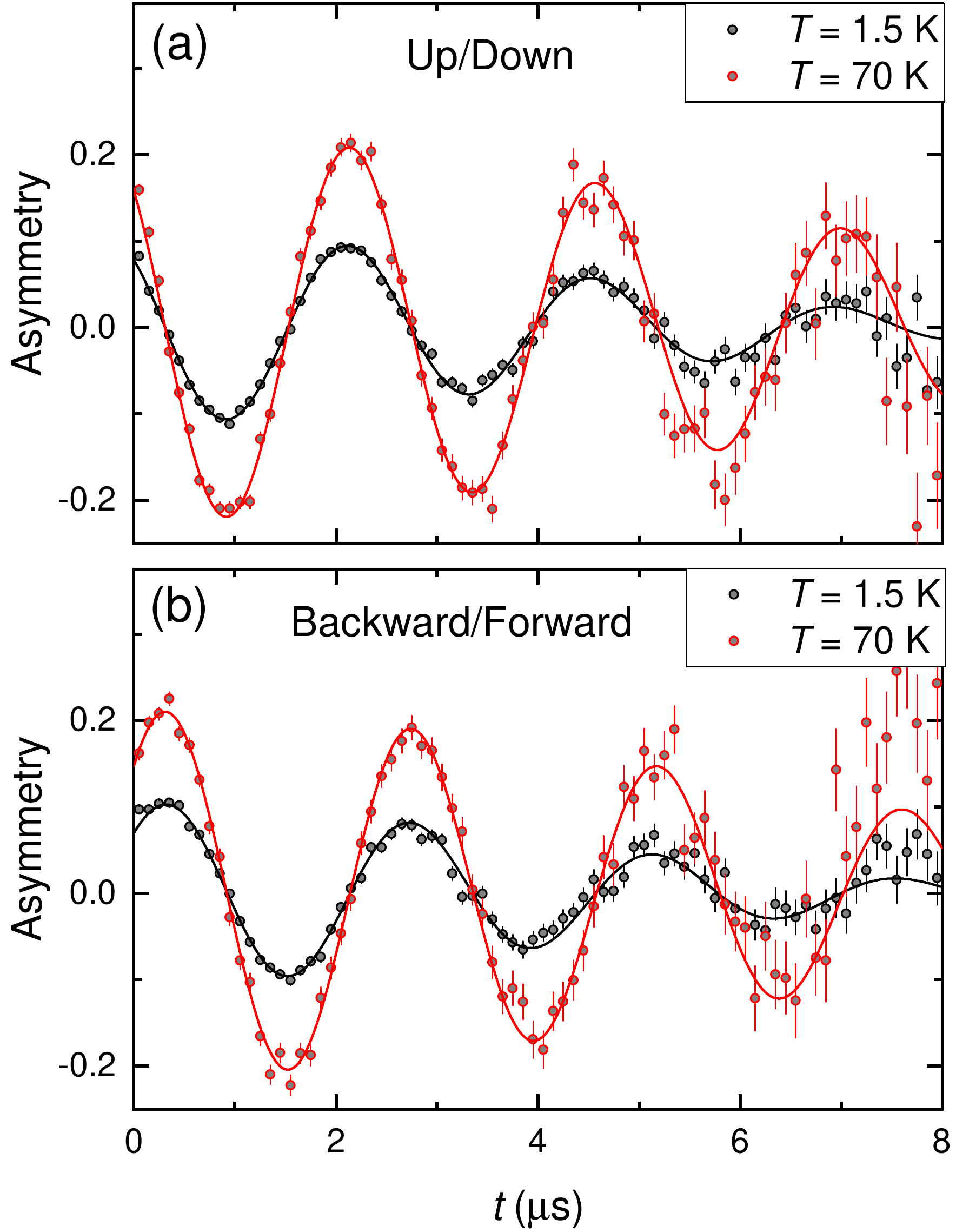}
% \vspace{-1.0cm}
%
\caption{  (a) WTF-$\mu$SR time-spectra ($B_{\rm ex}=3$~mT) of CaK(Fe$_{0.949}$Ni$_{0.051}$)$_4$As$_4$ measured on Up/Down set of positron counters below ($T\simeq1.5$~K) and above ($T=70$~K) the magnetic transition temperature ($T_N\simeq50$~K). The solid lines are fits of Eq.~\ref{eq:Asymmetry-WTF} to the data (see text for details). (b) The same as is panel (a) but for Backward/Forward set of detectors. }
 \label{fig:WTF-signal}
\end{figure}

Figure~\ref{fig:WTF-signal} shows the wTF-$\mu$SR time-spectra of CaK(Fe$_{0.949}$Ni$_{0.051}$)$_4$As$_4$ single crystal sample measured above ($T\simeq70$~K) and below ($T\simeq1.5$~K) the magnetic transition ($T_N\simeq50$~K). The external magnetic field $B_{\rm ex}=3$~mT was applied
along the $x-$direction [see Fig.~\ref{fig:muSR_setup}~(a)]. The angle between the initial muon-spin polarization and the muon momentum was set to 45$^{\rm o}$. The panels (a) and (b) of Fig.~\ref{fig:WTF-signal} correspond to the data collected on Up/Down and Backward/Forward set of detectors.

The time evolution of the muon-spin asymmetry in $\mu$SR experiments $A(t)$ can be described as:
\begin{equation}
A(t)=A(0)P(t)= A_{\rm s}(0) P_{\rm s}(t)+A_{\rm bg}(0) P_{\rm bg}(t).
 \label{eq:Asymmetry-general}
\end{equation}
Here the indexes 's' and 'bg' denote the sample and background contributions, respectively. $A_{\rm s}(0)$[$A_{\rm bg}(0)$] are the initial asymmetry and $P_{\rm s}(t)$[$P_{\rm bg}(t)$] the time evolution of the muon-spin polarization belonging to the sample(background).  The background component represent the muons missing the sample and stopped, {\it e.g.}, in Ag degrader sheets [see Fig.~\ref{fig:muSR_setup}~(b)], cryostat walls, cryostat windows, {\it etc.}
Considering the transition of CaK(Fe$_{0.949}$Ni$_{0.051}$)$_4$As$_4$ from non magnetic (nm) to the magnetic (m) state,\cite{Meier_NPJ_2018, Meier_PhD-Thesis_2018, Kreyssig_PRB_2018, Budko_PRB_2018} the sample contribution was further assumed to consist of two parts:\cite{Khasanov_CrAs-Scirep_2015}
\begin{equation}
A_{\rm s}(t) = A_{\rm nm}(0)P_{\rm nm}(t) +A_{\rm m}(0)P_{\rm m}(t).
\label{eq:Asymmetry-sample}
\end{equation}

In analysis of wTF-$\mu$SR data only the non magnetic sample component was considered. The magnetic term [$A_{\rm m}(0)P_{\rm m}(t)$] vanishes within the first $\sim 0.1$~$\mu$s (see Sec.~\ref{seq:ZF-LT}) and thus it is not observed with the present data binning ($\simeq 0.126$~$\mu$s). The 'nm', and 'bg' contributions in Eqs.~\ref{eq:Asymmetry-general} and \ref{eq:Asymmetry-sample} were further combined into the single term:
\begin{equation}
A(t)=A(0)\cos (\gamma_\mu B_{ex}t+\phi)\  e^{-\sigma^2t^2/2}.
 \label{eq:Asymmetry-WTF}
\end{equation}
Here $\phi$ is the initial phase of the muon-spin ensemble, and $\sigma$ is the Gaussian relaxation rate. The solid lines in Fig.~\ref{fig:WTF-signal} correspond to the fit of Eq.~\ref{eq:Asymmetry-WTF} to the wTF-$\mu$SR data.

\begin{figure}[htb]
\centering
\includegraphics[width=0.9\linewidth]{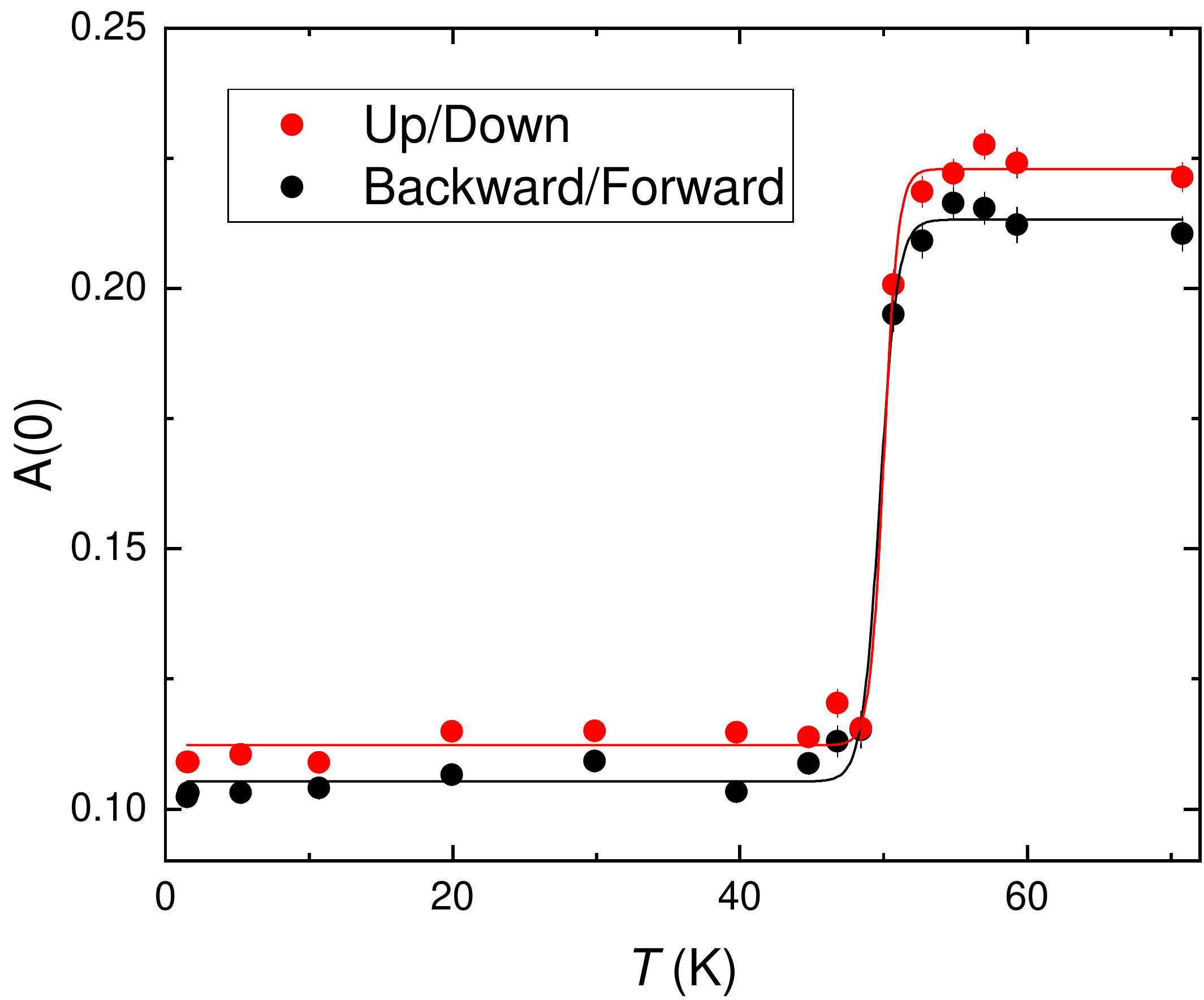}
% \vspace{-1.0cm}
%
\caption{  Temperature dependencies of the initial asymmetry $A(0)$ for the Up/Down and Backward/Forward set of positron counters. The solid lines are fits of Eq.~\ref{eq:Asymmetry-WTF-fit} to $A(0)$ {\it vs.} $T$ data. See text for details.}
 \label{fig:WTF-A0}
\end{figure}

Figure \ref{fig:WTF-A0} shows the temperature dependence of the initial asymmetry $A(0)$ obtained from the fit of wTF-$\mu$SR data by means of Eq.~\ref{eq:Asymmetry-WTF}. The magnetic ordering temperature $T_{\rm N}$ and the width of the magnetic transition $\Delta T_{\rm N}$ were determined by using the phenomenological function:\cite{Khasanov_OIE_PRL_2008}
\begin{equation}
A(0,T) = A_{\rm s}(0)\cdot\frac{1}{1+\exp([T_{\rm N}-T]/\Delta T_{\rm N})} + A_{\rm bg}(0).
 \label{eq:Asymmetry-WTF-fit}
\end{equation}
The results of the fit are represented by solid lines. The fit results for the Up/Down [Backward/Forward] set of detectors are: $T_{\rm N} = 50.0(2)~K [49.7(2)$~K], $\Delta T=0.5(1)$~K [0.6(1)~K], $A_{\rm s}(0) =0.111(2)$ [0.108(2)] , and $A_{\rm bg}(0)=0.112(1)$ [0.105(1)].

The results of wTF-$\mu$SR experiments can be summarised as follows:\\
 (i) The value of the magnetic ordering temperature $T_{\rm N} = 49.9(3)$~K coincides rather well with 50.0(6)~K obtained by Kreyssig {\it et al.}\cite{Kreyssig_PRB_2018} in resistivity, specific heat, and neutron scattering experiments on the sample with the similar doping level. \\
 (ii) The width of the transition $\Delta T_{\rm N} = 0.5(2)$~K is rather small suggesting that the magnetic order sets inside the sample uniformly. In other words, the magnetic ordering temperature $T_{\rm N}$ stays the same (within $\simeq 0.5$~K accuracy) over the full sample volume. \\
 (iii) The fact that $A_{\rm s}(0)\simeq A_{\rm bg}(0)$ implies that 50\% of all the muons stop in the sample, while the rest contribute to the background.

\subsection{Zero-field $\mu$SR experiments}

\subsubsection{$P^{\perp c}$ and $P^{\parallel c}$ set of data \label{seq:ZF-LT}}

\begin{figure*}[htb]
\centering
\includegraphics[width=1.0\linewidth]{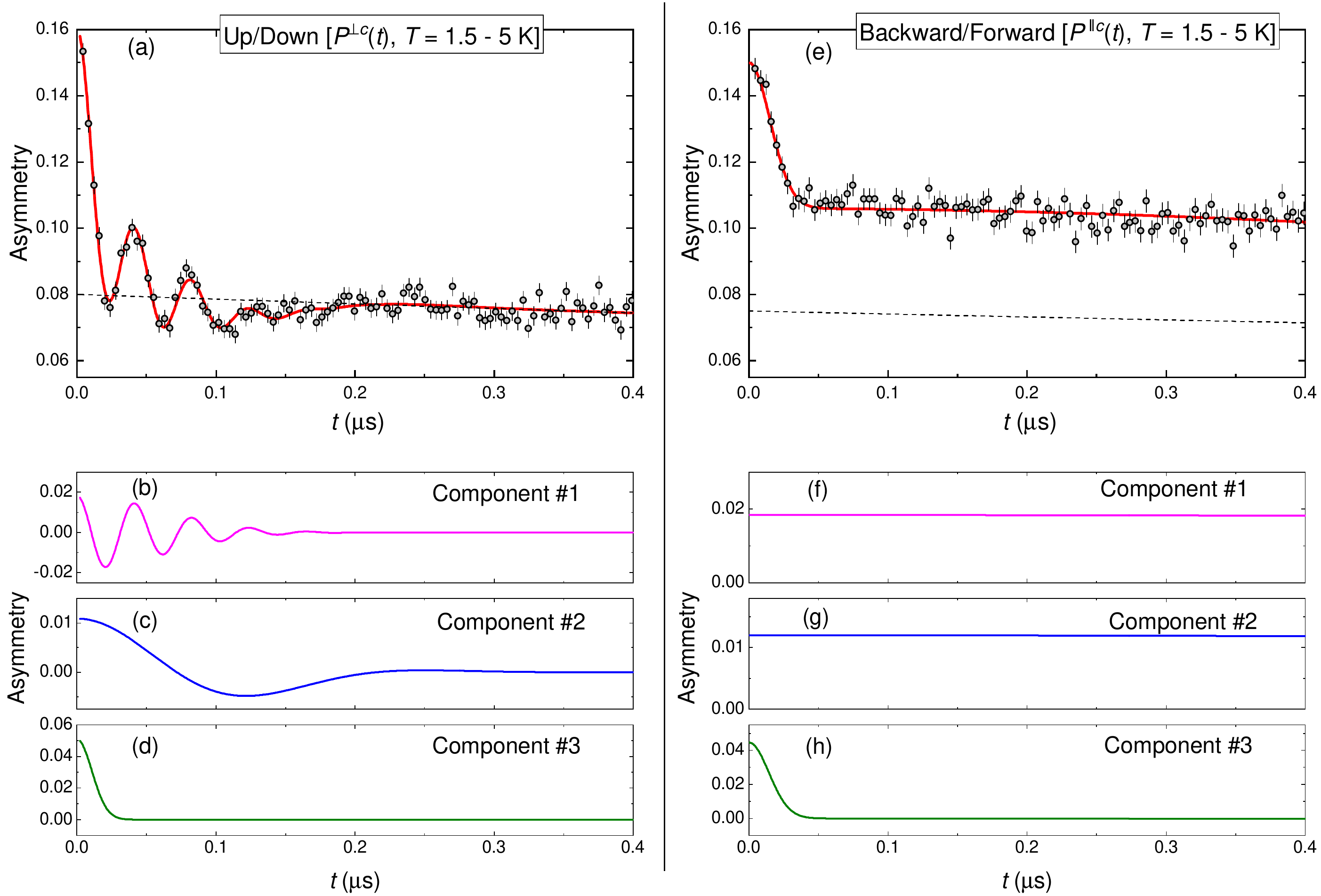}
% \vspace{-1.0cm}
%
\caption{(a) ZF-$\mu$SR time-spectra of CaK(Fe$_{0.949}$Ni$_{0.051}$)$_4$As$_4$ measured on Up/Down set of positron counters. The experiment probes the time evolution of $P^{\perp c}$ component of the muon-spin polarization. In order to improve statistics, the data sets collected at temperatures ranging from 1.5K up to 5~K are combined together. The solid line is the fit of Eq.~\ref{eq:Asymmetry-general} with the sample part described by Eq.~\ref{eq:Asymmetry-ZF} to the data. The dashed line is the time evolution of the background component. (b), (c), and (d) are  contributions of components \#1, \#2, and \#3, respectively. (e) is the same as in panel (a), but for Backward/Forward set of positron counters. In this experiment the time evolution of $P^{\parallel c}$ component of the muon-spin polarization is probed.  The panels (f), (g), and (h) represent contributions of the first, second and the third components, respectively. }
 \label{fig:ZF_spectra}
\end{figure*}

In ZF-$\mu$SR experiments, the muon-spin precesses in internal field(s) at the muon stopping site(s) which are created by the surrounding magnetic moments  (nuclear or electronic in origin).
Figure~\ref{fig:ZF_spectra} shows the ZF-$\mu$SR time-spectra collected on  Up/Down (panel a) and Backward/Forward (panel e) set of positron counters. In order to improve statistics, the data sets collected at temperatures ranging from $\simeq 1.5$ up to 5~K are combined together. Obviously, oscillations of the muon-spin polarization corresponding to the precession of the muon-spin in internal field ($B_{\rm int}$) are observed for the Up/Down, but they are missing  for the Backward/Forward set of detectors. Bearing in mind that the Up/Down and Backward/Forward responses correspond to the time evolution of $P^{\perp c}$ and $P^{\parallel c}$ components of the muon-spin polarization, respectively (see Sec.~\ref{seq:muSR} and Fig.~\ref{fig:muSR_setup}), one concludes that internal fields on the muon stopping sites are aligned along the $c-$axis of the CaK(Fe$_{0.949}$Ni$_{0.051}$)$_4$As$_4$ single crystal.

The analysis of the sample response in ZF-$\mu$SR experiments was performed by considering the presence of three muon stopping sites as inferred from the muon-site calculations (see Seq.~\ref{seq:stopping-sites} and Fig.~\ref{fig:crystal-structure}):
\begin{equation}
A_{\rm m}^{\rm ZF}(t) = \sum_{i=1}^3 \; A_{i}(0) \; e^{-\sigma_i^2 t^2/2} \; \cos(\gamma_{\mu} B_{{\rm int},i}t).
 \label{eq:Asymmetry-ZF}
\end{equation}
Here $A_i(0)$, $\sigma_i$, and $B_{{\rm int},i}$ are the initial asymmetry, the Gaussian relaxation rate, and the internal field of the $i-$th component, respectively.
Fits of Eq.~\ref{eq:Asymmetry-general} with the sample part described by Eq.~\ref{eq:Asymmetry-ZF} to the ZF-$\mu$SR data are presented in Figs.~\ref{fig:ZF_spectra}~(a) and (e) by solid red lines. The parameters obtained from the fits are summarized in Table~\ref{tab:table1}.

From the results of the fit the following three important points emerge: \\
 (i) Three components (two oscillating, \#1 and \#2, and one fast relaxing non-oscillating, \#3) are clearly resolved by fitting the Up/Down ($P^{\perp c}$) set of data. The time evolutions of these components are presented in panels (b), (c), and (d) of Fig.~\ref{fig:ZF_spectra}. \\
 (ii) The fit of the Backward/Forward ($P^{\parallel c}$) set of data could be performed by using only two components (the slow and the fast relaxing ones). The relative weight (the fraction) of the fast relaxing component is the {\it same} (within the experimental accuracy) as the non-oscillating fast relaxing one in the Up/Down ($P^{\perp c}$) set of data.  It is reasonable to assume, therefore, that in both set of experiments the non-oscillating fast relaxing component (\#3) originates from the same muon stopping site. The slow relaxing contribution was further assigned to two oscillating components (\#1 and \#2) observed in Up/Down set of positron counters.\\
 (iii) The ratio between the sample and the background asymmetries was fixed to that determined in wTF experiments [$A_{\rm s}(0)\simeq A_{\rm bg}(0)\simeq0.5 A(0)$]. The time evolution of the background component is presented in Figs.~\ref{fig:ZF_spectra}~(a) and (e) by dashed lines.

\begin{table*}[htb]
\caption{\label{tab:table1} Parameters obtained from the fit of ZF-$\mu$SR time-spectra of CaK(Fe$_{0.949}$Ni$_{0.051}$)$_4$As$_4$. The meaning of the parameters is: $B_{{\rm int},i}$ is the internal field, $f_i$ is the volume fraction [$f_i = A_i(0)/\{A_1(0)+A_2(0)+A_3(0)\}$], and $\sigma_i$ is the Gaussian relaxation rate of the $i-$th component, respectively. }
\begin{tabular}{c|c|ccc|ccc|ccc}
\hline
\hline
Polarisation&Detector set&\multicolumn{3}{c|}{$B_{\rm int}$~(mT)}& \multicolumn{3}{c|}{Volume Fraction}&\multicolumn{3}{c}{Relaxation rate ($\mu{\rm s}^{-1}$)}\\
component&&$B_{{\rm int},1}$&$B_{{\rm int},2}$&$B_{{\rm int},3}$&$f_1$&$f_2$&$f_3$&$\sigma_1$&$\sigma_2$&$\sigma_3$\\
\hline
$P^{\perp c}$&Up/Down& 177.2(1.5)&25.7(1.3)&0 &0.23(2)&0.15(1)&0.62(6)&16.3(1.4)& 9.7(1.2) &92.8(7.4)\\
$P^{\parallel  c}$& Backward/Forward&0&0&0& 0.25\footnotemark[1]&0.16\footnotemark[1]&0.59(7)&0.39(6)& 0.39(6)& 69.9(5.3)\\
\hline
\hline
\footnotetext[1]{The ratio between $f_1$ and $f_2$ was kept the same as for $P^{\perp c}$ set of experiments}
\end{tabular}
\end{table*}

\subsubsection{Temperature dependence of the internal field $B_{\rm int}$}

The value of the internal field at the muon stopping position is determined by surrounding magnetic moments. In a case of magnetically ordered samples $B_{\rm int}$ is directly proportional to the value of the ordered magnetic moments (Fe moments in a case of CaK(Fe$_{0.949}$Ni$_{0.051}$)$_4$As$_4$ studied here). Consequently, the temperature dependence of $B_{\rm int}$ reflects precisely the temperature evolution of the magnetic order parameter $B_{\rm int}\propto m_{\rm Fe}$.

\begin{figure}[htb]
\centering
\includegraphics[width=0.95\linewidth]{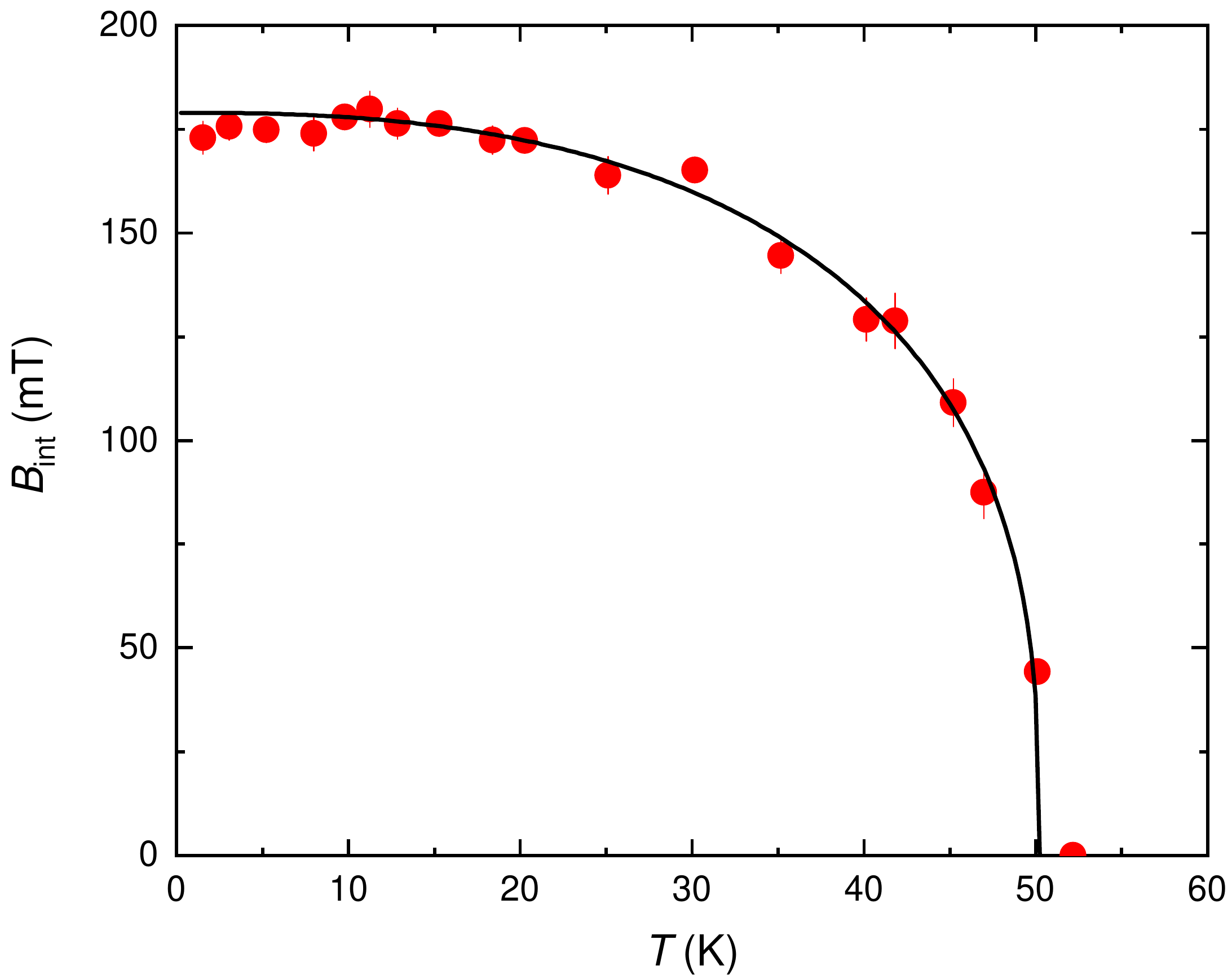}
% \vspace{-1.0cm}
%
\caption{Temperature dependence of the internal field $B_{{\rm int},1}$. The solid line is the fit of Eq.~\ref{eq:power-law} to the data with the parameters $T_{\rm N} = 50.2(3)$~K, $\alpha=2.52(9)$, $\beta=0.35(2)$, and $B_{\rm int}(0)= 179(1)$~mT. }
 \label{fig:Bint}
\end{figure}

The temperature dependence of $B_{{\rm int},1}$ is shown in Fig.~\ref{fig:Bint}. The fit was performed globally. In this case the Eq.~\ref{eq:Asymmetry-general} was fit to the full set of ZF Up/Down data with certain parameters kept global and some of them remaining individual for each particular data set. The 'global' parameters were the initial asymmetries [$A_{1}(0)$, $A_2(0)$, $A_3(0)$] and the ratio between the internal fields ($B_{{\rm int},2}/ B_{{\rm int},1}$). The individual parameters were $B_{{\rm int},1}$, and the relaxation rates $\sigma_1$, and $\sigma_3$. Note that the preliminary fit with all the parameters remaining 'free' reveal that the relaxation rates $\sigma_1$ and $\sigma_2$ stay almost equal. So it was assumed, additionally,  $\sigma_1=\sigma_2$. Above the magnetic transition, a weak Gauss-Kubo-Toyabe damping of the signal was observed caused by the dipole-dipole interaction
of the muon magnetic moment with randomly oriented nuclear magnetic moments in the paramagnetic temperature regime.

The results presented in Figure~\ref{fig:Bint} imply that the onset temperature of the long-range magnetic order, corresponding to decreasing $B_{\rm int}$ down to zero, is consistent with the results of resistivity and the neutron scattering experiments of Kreyssig {\it et al.},\cite{Kreyssig_PRB_2018} as well as with the results of wTF-$\mu$SR measurements presented in Sec.~\ref{seq:wTF}. The temperature dependence of $B_{\rm int}$ is characteristic of a second-order phase transition, consistent with the results of heat capacity, $^{75}$As nuclear magnetic resonance (NMR), neutron scattering, and M\"{o}ssbauer spectroscopy experiments.\cite{Meier_NPJ_2018, Meier_PhD-Thesis_2018, Kreyssig_PRB_2018, Budko_PRB_2018}

The temperature dependence of $B_{\rm int}(T)$ was analyzed using a fit to the temperature-dependent magnetic order parameter of the form:\cite{Pratt_JPCM_2007}
\begin{equation}
B_{\rm int}(T)=B_{\rm int}(0) \left[ 1- \left( \frac{T}{T_{\rm N}} \right)^\alpha \right]^\beta.
 \label{eq:power-law}
\end{equation}
The fit was made by considering points above the superconducting transition temperature $T_{\rm c}\simeq 9$~K. The small reduction of $B_{\rm int}$ below $T_{\rm c}$ is caused by interaction between the magnetic and superconducting order parameters and it is discussed later. The fit yields $T_{\rm N} = 50.2(3)$~K, $\alpha=2.52(9)$, $\beta=0.35(2)$, and $B_{\rm int}(0)= 179(1)$~mT. The value of the effective critical exponent $\beta$ lies quite close to the critical exponent $\beta =0.325$ expected for a 3D magnetic system (3D
Ising universality class).\cite{Jongh_AdPh_1974}

\section{Discussion \label{seq:discussion}}

\subsection{Consistency of ZF-$\mu$SR results with the 'hedgehog'-type of magnetic order}

The results of Sec.~\ref{seq:ZF-LT} reveal the presence of 3 contributions to the time evolution of the muon-spin polarization (see components \#1,  \#2 and \#3 in Fig.~\ref{fig:ZF_spectra} and Table~\ref{tab:table1}). Considering the results of muon-site calculations presented in Sec.~\ref{seq:stopping-sites}, these 3 components could be further assigned to 3 different muon stopping sites within the unit cell of CaK(Fe$_{1-x}$Ni$_{x}$)$_4$As$_4$. Internal fields at 2 muon stopping positions (components \#1 and \#2) are aligned along the crystallographic $c-$direction, while the very broad distribution of fields with the average value centered at zero (component \#3) corresponds to the third muon position.

The symmetry analysis calculations presented in Appendix~\ref{sec:Symmetry_Analysis} considers eight different magnetic spin-vortex-crystal (SVC) structures with orthogonal iron moments lying in the $ab-$plane. The structures were presented by magnetic order parameters from one dimensional irreducible representations $\tau 1 - \tau 8$. The SVC structures were divided into two groups with so called 'hedgehog' and 'loop' motif in accordance with its orthogonal arrangement of the spin pattern.\cite{Meier_NPJ_2018} The SVC structures preserve the C4 symmetry and become consistent with the tetragonal lattice symmetry.

\begin{figure}[htb]
\centering
\includegraphics[width=0.7\linewidth]{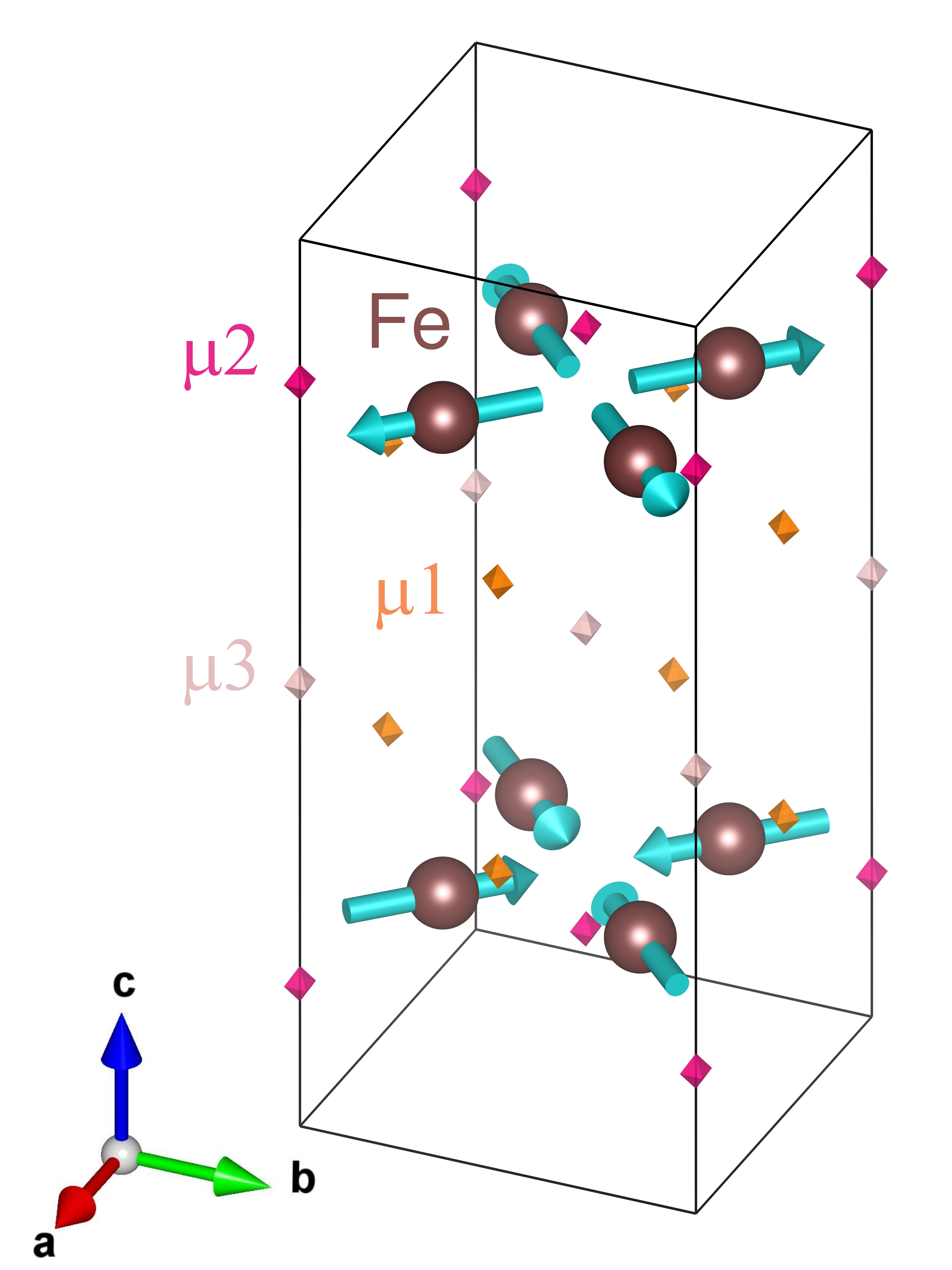}
% \vspace{-1.0cm}
%
\caption{  The magnetic structure of  CaK(Fe$_{1-x}$Ni$_{x}$)$_4$As$_4$ which is consistent with the results of ZF-$\mu$SR measurements. Only the magnetic (Fe) ions are shown. Arrows represent the ordered Fe moments. Diamond symbols correspond to different muon stopping sites. The magnetic unit cell ($P4/mbm$ group representation) is doubled and 45$^{\rm o}$ rotated in the $ab-$plane with respect to the crystallographic one ($P4/mmm$ group representation). See Appendix~\ref{sec:Symmetry_Analysis} for details. The structure is visualized by using VESTA package, Ref.~\onlinecite{Vesta}.}
 \label{fig:magnetic-structure}
\end{figure}

The results of Appendix~\ref{sec:Symmetry_Analysis} exclude the 'loop' SVC structures from consideration. With 4 ’hedgehog’ SVC  magnetic structures left, the one, corresponding to the $\tau1$ irreducible representation, results in two different fields at the muon stopping sites $\mu2$ and $\mu3$ and the zero-field on the site $\mu1$, respectively. The rest gives the single field either at the site $\mu2$ ($\tau_4$) or $\mu1$ ($\tau_5$ and $\tau_8$) and zero-fields at the muon sites left (see Table~\ref{tab:table2} in Appendix~\ref{sec:Symmetry_Analysis}). In all cases only $z$ component is present, which corresponds to an alignment of the internal magnetic field on the muon position along the crystallographic $c-$direction.

By comparing these results with the ZF-$\mu$SR data (see Fig.~\ref{fig:ZF_spectra} and Table~\ref{tab:table1}), we conclude that only one single type of the magnetic structure becomes consistent with the experiment. This is the SVC structure with the 'hedgehog' motif corresponding to the $\tau1$ irreducible representation. The arrangement of Fe moments is presented in Fig.~\ref{fig:magnetic-structure}. This structure is characterized by noncollinear Fe moments featuring an alternating all-in and all-out motif around the As1 sites within the $ab-$planes and antiferromagnetic coupling along the $c-$direction. Overall, this structure is fully consistent with that reported in NMR, M\"{o}ssbauer, and neutron scattering experiments.\cite{Meier_NPJ_2018, Kreyssig_PRB_2018}

Following results presented in Appendix~\ref{sec:Symmetry_Analysis}, the component \#1, \#2, and \#3 of ZF-$\mu$SR signal (see Sec.~\ref{seq:ZF-LT} and Fig.~\ref{fig:ZF_spectra}) could be assigned to $\mu2$, $\mu3$ and $\mu1$ muon-stopping sites, respectively. Three important points need to be considered:\\
(i) The calculations predict the dipolar fields at the first and the second muon stopping sites resulting in $B_{{\rm int}, \mu2}\simeq366$~mT per 1~$\mu_{\rm B}$ and $B_{{\rm int}, \mu 3}\simeq94$~mT per 1~$\mu_{\rm B}$, respectively ($\mu_{\rm B}$ is the Bohr magneton). With the experimentally measured $B_{{\rm int},1}\simeq 177.2$~mT and $B_{{\rm int},2}\simeq 25.7$~mT (see Table~\ref{tab:table1}) the value of the ordered Fe moments is found to be $m_{\rm Fe}=0.38(0.11)$~$\mu_{\rm B}$. Such value stays in agreement with $m_{\rm Fe}=0.37(10)$~$\mu_{\rm B}$ obtained in neutron scattering experiments.\cite{Kreyssig_PRB_2018}\\
(ii) The fact that the values of the magnetic moment obtained from $B_{{\rm int},1}$ ($m_{\rm Fe}\simeq 0.48$~$\mu_{\rm B}$) and $B_{{\rm int},2}$ ($m_{\rm Fe}\simeq 0.27$~$\mu_{\rm B}$) are $\simeq 45$\% different could be explained by taking into account that calculations consider only the local dipolar fields and neglect contact hyperfine contributions. Similar differences were found {\it e.g.} in La$_2$CuO$_4$,\cite{Klauss_JPCM_2004} $R$FeAsO ($R=$La, Ce, Pr, and Sm),\cite{Maeter_PRB_2009} Ba$_{1-x}$K$_x$Fe$_2$As$_2$,\cite{Mallett_EPL_2015}  Ba$_{1-x}$Na$_x$Fe$_2$As$_2$,\cite{Sheveleva_Arxiv_2020} MnP,\cite{Khasanov_MnP_PRB_2016, Khasanov_MnP_JPCM_2017} {\it etc.}\\
(iii) The occupancy of the muon stopping sites predicted by calculations is: $\mu1 \rightarrow \mu2 \rightarrow \mu3$. Experiment reveals the volume fractions of the first, second, and third component of ZF-$\mu$SR signal to follow: $f_3 \rightarrow f_1 \rightarrow f_2$ (see Table~\ref{tab:table1}). Considering the assignment of ZF-$\mu$SR signal components to the muon stopping sites (see above), this leads to an additional agreement between the theory and the experiment.

\subsection{Coexistence of superconductivity and magnetism in CaK(Fe$_{0.949}$Ni$_{0.051}$)$_4$As$_4$}

Theory works of Fernandes {\it et al.},\cite{Fernandes_PRB_2010} Vorontsov {\it et al.},\cite{Vorontsov_PRB_2010} and Schmiedt {\it et al.}\cite{Schmiedt_PRB_2014} reveal that magnetism in Fe-based superconductors may coexist with superconductivity. As a result, a commensurate spin-density wave (SDW) can coexist with a superconducting $s\pm$ state. In the case of coexistence of magnetic order and superconductivity, an interaction between both order parameters is expected. This may change the magnitude of the order parameters and alter the critical temperatures with respect to the decoupled situation.

Here we used the approach of Machida,\cite{Machida_JPSJ_1981} who has considered the coexistence of spin-density wave (SDW) type of magnetism with superconductivity within the three dimensional single band case. The model was recently employed by Bud'ko {\it et al.} \cite{Budko_PRB_2018} with the possible application to the multiple-band materials as, {\it e.g.} CaK(Fe$_{0.949}$Ni$_{0.051}$)$_4$As$_4$ studied here. Within this model the SDW order is assumed to develop over a nested
part of the Fermi surface whereas the superconductivity forms over the full Fermi surface(s).

\subsubsection{Temperature dependence of the magnetic order parameter \label{seq:T-dependece_of_m}}

\begin{figure}[htb]
\centering
\includegraphics[width=0.95\linewidth]{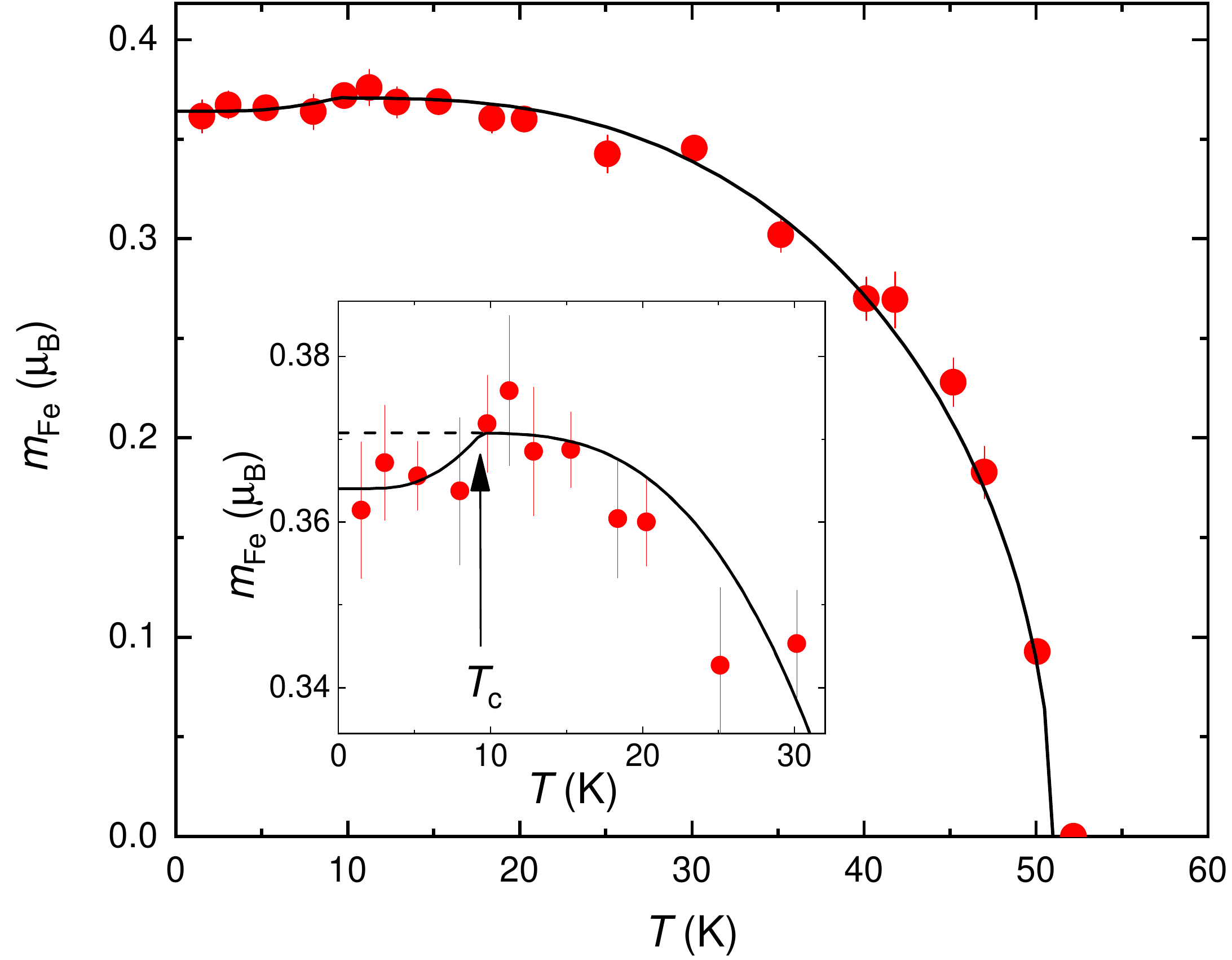}
% \vspace{-1.0cm}
%
\caption{ Temperature dependence of the magnetic order parameter $m_{\rm Fe}$ of CaK(Fe$_{0.949}$Ni$_{0.051}$)$_4$As$_4$. The solid line is the analysis by using the approach of Machida.\cite{Machida_JPSJ_1981, Budko_PRB_2018} The inset shows extension of the low-temperature part. The position of the superconducting transition temperature $T_{\rm c}$ is marked by the arrow. }
 \label{fig:mFe}
\end{figure}

Following Refs.~\onlinecite{Budko_PRB_2018} and \onlinecite{Machida_JPSJ_1981} the superconducting ($\Delta$) and the magnetic ($M$) order parameters could be obtained by solving the system of two coupled self-consistent equations:
%
%\begin{widetext}
\begin{eqnarray}
  \ln\frac{T}{T_{\rm s0}} &=& 2\pi T \sum_{\omega>0}^{\omega_{\rm s}}   \biggl[ \frac{1}{2M} \biggl( \frac{M+\Delta}{\sqrt{\omega^2+(M+\Delta)^2}} \nonumber \\ &&+\frac{M-\Delta}{\sqrt{\omega^2+(M-\Delta)^2}}   \biggl) - \frac{1}{\omega} \biggl]
  \label{eq:Machida1}
\end{eqnarray}
and
\begin{eqnarray}
  \ln\frac{T}{T_{\rm c0}} &=& n_1 2\pi T \sum_{\omega>0}^{\omega_{\rm D}}   \biggl[ \frac{1}{2\Delta} \biggl( \frac{\Delta+M}{\sqrt{\omega^2+(\Delta+M)^2}} \nonumber \\ &&+\frac{\Delta-M}{\sqrt{\omega^2+(\Delta-M)^2}}   \biggl) - \frac{1}{\omega}  \biggl] \nonumber \\
  && + n_2 2\pi T \sum_{\omega>0}^{\omega_{\rm D}}   \biggl(\frac{1}{\sqrt{\omega^2+\Delta^2}}-\frac{1}{\omega} \biggl).
  \label{eq:Machida2}
\end{eqnarray}
%\end{widetext}
Here, $\omega = \pi T(2n + 1)$ are Matsubara frequencies with a positive integer $n$, $\omega_{\rm D}$ is the Debye frequency, $\omega_{\rm s}$ is a corresponding limit for SDW, $T_{\rm s0}$ is the SDW ordering temperature,  $T_{\rm c0}$ is the superconducting transition temperature in absence of magnetism, $n_1$ is partial density of states (DOS) on the Fermi surface part responsible for SDW, and $n_2=1-n_1$. The sums in Eqs.~\ref{eq:Machida1} and \ref{eq:Machida2} are convergent and for $\omega_{\rm D}\gg T_{\rm c0}$ and $\omega_{\rm s}\gg T_{\rm s0}$ the upper limits of summation
can be extended to infinity. Only the case when first the magnetic and then the superconducting order sets in, {\it i.e.} for $T_{\rm s0}>T_{\rm c0}$, is considered.

The comparison of Machida's approach with the temperature evolution of the magnetic order parameter obtained in the present study is shown in Fig.~\ref{fig:mFe}. The $m_{\rm Fe}(T)$ dependence was obtained by normalizing the $B_{\rm int}(T)$ curve (Fig.~\ref{fig:Bint}) to $m_{\rm Fe}(20$\rm ~K$)=0.37$~$\mu_{\rm B}$ as obtained by Kreyssig {\it et al.}\cite{Kreyssig_PRB_2018} in neutron scattering experiments. The parameters of the modelled curve are: $T_{\rm s0}/T_{\rm c0}=2$ and $n_1=0.35$.
Note that the theory captures all major features of the experimentally obtained $m_{\rm Fe}(T)$. The initial increase below $T_{\rm N}\simeq 51$~K, the saturation in $30\gtrsim T \gtrsim  10$~K temperature region and the slight drop below the superconducting transition temperature $T_{\rm c}\simeq 9$~K (see the inset in Fig.~\ref{fig:mFe}) are reproduced quite precisely.

It should be noted here that the theory of Machida does not allow to get a 'unique' values of the fitting parameters (see also the analysis of M\"{o}ssbauer spectroscopy data from Ref.~\onlinecite{Budko_PRB_2018}). Similar $m_{\rm Fe}(T)$ curves could be obtained with the different set of parameters.  As is stressed already in Ref.~\onlinecite{Budko_PRB_2018}, a unique
determination of fit parameters would require additional boundary conditions on them.

\subsubsection{Suppression of the magnetic order parameter in the superconducting state}

\begin{figure}[htb]
\centering
\includegraphics[width=0.95\linewidth]{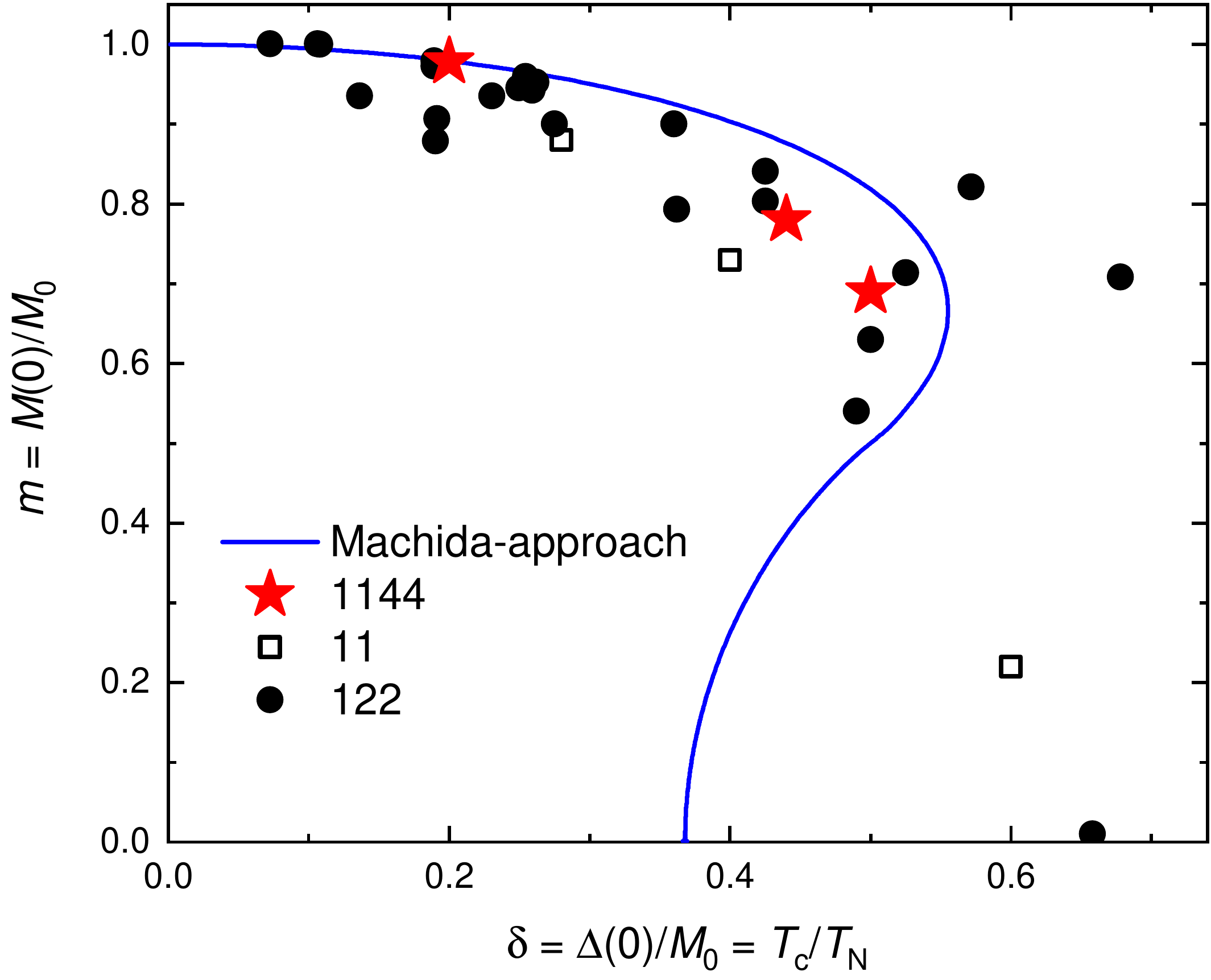}
% \vspace{-1.0cm}
%
\caption{ Dependence of the normalized magnetic $m =M(0)/M_0$ {\it vs.} the superconducting order parameter $\delta=\Delta(0)/M_0$ obtained within the framework of Machida approach,\cite{Machida_JPSJ_1981} [$M_0$ is the zero-temperature value the magnetic order parameter in absence of superconductivity, while  $M(0)=M(T=0)$ and $\Delta(0)=\Delta(T=0)$ are $T=0$ values of the magnetic and the superconducting order parameters, respectively].
The red stars are experimental data obtained within present studies and by means of M\"{o}ssbauer spectroscopy and neutron scattering experiments in CaK(Fe$_{1-x}$Ni$_{x}$)$_4$As$_4$ (1144 family of Fe-SC's).\cite{Budko_PRB_2018, Kreyssig_PRB_2018} The open squares and closed circles are the experimental data for samples belonging to  11,\cite{Bendele_PRL_2010, Bendele_PRB_2010} and 122 families of Fe-SC's,\cite{Goldman_PRB_2008, Avci_PRB_2011, Kim_PRB_2011, Kreyssig_PRB_2010, Wang_PRL_2012, Luo_PRL_2012, Nandi_PRL_2010, Marsik_PRL_2010, Christianson_PRL_2009, Pratt_PRL_2011, Bernhard_PRB_2012, Materne_PRB_2015, Sheveleva_Arxiv_2020, Li_PRB_2012, Zhou_NatCom_2013} respectively. }
 \label{fig:m_vs_Tc}
\end{figure}

The theory of Machida allow to obtain the reduction of the magnetic order parameter in the presence of superconductivity.
Calculations presented in Appendix~\ref{sec:Machida-Approach} show that at the limit of $T \rightarrow 0$, Eqs.~\ref{eq:Machida1} and \ref{eq:Machida2} transforms to:
\begin{equation}
(m+\delta)\ln(m+\delta) + (m-\delta)\ln(|m-\delta|) = 0
 \label{eq:Machida_reduced1}
\end{equation}
and
\begin{equation}
\delta\ln(  \delta \; T_{\rm s0}/T_{\rm c0}) +n_1 [(m+\delta)\ln(m+\delta)-d\ln \delta]=0.
  \label{eq:Machida_reduced2}
\end{equation}
Here $\delta$ and $m$ are dimensionless superconducting and magnetic order parameters defined as: $\delta = \Delta(T=0)/M_0$ and  $m=M(T=0)/M_0$ ($M_0$ is the zero-temperature value of the magnetic order parameter in the absence of superconducting order).
Solutions for $m$ {\it vs.} $\delta$ values follow the blue solid line presented in Figure~\ref{fig:m_vs_Tc}. The upper part of the curve corresponds to the monotonic decrease of $m$ with increasing $\delta$. By approaching the point $\delta\simeq 0.55$, $m\simeq 0.67$, the tendency changes. With further decreasing $m$, $\delta$ decreases as well, by approaching $\delta\simeq 0.37$ at $m=0$. The theory also predicts that $\delta$ remains always smaller than 1 and that the maximum value it may achieve corresponds to $\delta\simeq 0.55$ at $m\simeq 0.7$. This suggests that the magnetism is more robust than superconductivity and that the presence of nearly negligible magnetic contribution (see the point $\delta\simeq 0.37$ for $m \rightarrow 0$) is enough to suppress partially the superconductivity .

Comparison with the experimental data could be made by taking into account that within the Machida's approach:
\begin{equation}
  \delta=\frac{\Delta(T=0)}{M_0}=\frac{T_{\rm c}}{T_{\rm s0}}=\frac{T_{\rm c}}{T_{\rm N}}
\end{equation}
The closed and open symbols in Fig.~\ref{fig:m_vs_Tc} correspond to the experimental data for various Fe-SC's belonging to 1144 [as CaK(Fe$_{0.949}$Ni$_{0.051}$)$_4$As$_4$],\cite{Budko_PRB_2018, Kreyssig_PRB_2018} 11,\cite{Bendele_PRL_2010, Bendele_PRB_2010} and 122 families.\cite{Goldman_PRB_2008, Avci_PRB_2011, Kim_PRB_2011, Kreyssig_PRB_2010, Wang_PRL_2012, Luo_PRL_2012, Nandi_PRL_2010, Marsik_PRL_2010, Christianson_PRL_2009, Pratt_PRL_2011, Bernhard_PRB_2012, Materne_PRB_2015, Sheveleva_Arxiv_2020, Li_PRB_2012, Zhou_NatCom_2013} Qualitatively, the experimental data follow the upper branch of $m$ {\it vs.} $\delta$ curve.
The question remains: whether or not the lower part of the curve ($m\lesssim 0.65$) is just a non-physical solution of the theory or it may correspond to something not yet experimentally observed. One could assume that varying certain external parameters, as {\it e.g.} pressure or density of impurities, may allow to verify all these features of Machida's model.

At the end of this Section we would mention that the reduction of the magnetic order parameter as a function of $T_c/T_{\rm N}$ for 122 family of Fe-SC's was studied by Materne {\it et al.}\cite{Materne_PRB_2015}  based on a Landau theory for coupled superconducting and magnetic order parameters. A quadratic relation between $M(0)/M_0$ and $T_c/T_{\rm N}$ was reported. Note, however, that the Landau approach used in Ref.~\onlinecite{Materne_PRB_2015} requires knowledge of 4 parameters.

\section{Conclusions \label{seq:conclusions}}

The zero-field (ZF) and weak transverse-field (wTF) muon-spin rotation/relaxation experiments on CaK(Fe$_{0.949}$Ni$_{0.051}$)$_4$As$_4$ single crystal sample were performed. The main results could be summarised as follows: \\
(i) The sharp transition to the magnetic state with the ordering temperature $T_{\rm N}=50.0(5)$~K  in both ZF and wTF-$\mu$SR experiments is detected. The value of $T_{\rm N}$ is found to be in agreement with 50.0(6)~K obtained in resistivity, specific heat, and neutron scattering experiments on samples with the similar doping level.\cite{Meier_NPJ_2018, Meier_PhD-Thesis_2018, Kreyssig_PRB_2018} \\
(ii) The calculation of the muon-stopping sites and the symmetry analysis allow to identify the type of the magnetic order in CaK(Fe$_{1-x}$Ni$_{x}$)$_4$As$_4$.  The long-range magnetic spin-vortex-crystal order with the hedgehog motif within the $ab-$plane and the antiferromagnetic stacking along the $c-$direction agrees with the experiment. The value of the ordered magnetic moment was estimated to be $m_{\rm Fe}=0.38(11)$~$\mu_{\rm B}$ in agreement with 0.37(10)~$\mu_{\rm B}$ from neutron scattering experiments.\cite{Kreyssig_PRB_2018} The type of the magnetic order is the same as determined in NMR, M\"{o}ssbauer  spectroscopy and neutron scattering experiments.\cite{Meier_NPJ_2018, Kreyssig_PRB_2018} \\
(iii) A reduction of the magnetic order parameter below the superconducting transition was detected. The temperature evolution of the magnetic order parameter $m_{\rm Fe}(T)$ is well reproduced within the approach of Machida,\cite{Machida_JPSJ_1981, Budko_PRB_2018} which accounts for coexistence of a spin-density wave magnetism and superconductivity. \\
(iv) The theory of Machida was further applied in order to follow the interplay/coexistence of the magnetic and superconducting order parameters for $T$ approaching zero. Comparison with the experiment reveals that the data points for various Fe-based superconducting materials belonging to 3 different families reproduces reasonably well only the upper branch of $M(0)/M_0$ {\it vs.} $T_{\rm c}/T_{\rm N}$ curve. The question, on whether or not the lower part of the curve [$M(0)/M_0\lesssim 0.65$] correspond to a non-physical solution or it has not yet been experimentally detected, remains unexplored.

\begin{acknowledgments}
This work was performed at the Swiss Muon Source (S$\mu$S), Paul Scherrer Institute (PSI, Switzerland). The authors acknowledge the technical support of Joel Verezhak. The work of GS was supported by the Swiss National Science Foundation (SNF-Grant No. 200021-175935 and SNF Mobility grant P2EZP2-178604). The work of WRM at Iowa State University was supported by the Gordon and Betty Moore Foundation's EPiQS Initiative through Grant GBMF4411. Work at Ames Laboratory was supported by the U.S. Department of Energy, Office of Science, Basic Energy Sciences, Materials Science and Engineering Division. Ames Laboratory is operated for the U.S. DOE by Iowa State University under Contract No. DE-AC02-07CH11358.
\end{acknowledgments}

\appendix
\section{Dipolar fields at the muon stopping sites \label{sec:Symmetry_Analysis}}

\begin{figure}[htb]
\centering
\includegraphics[width=0.8\linewidth]{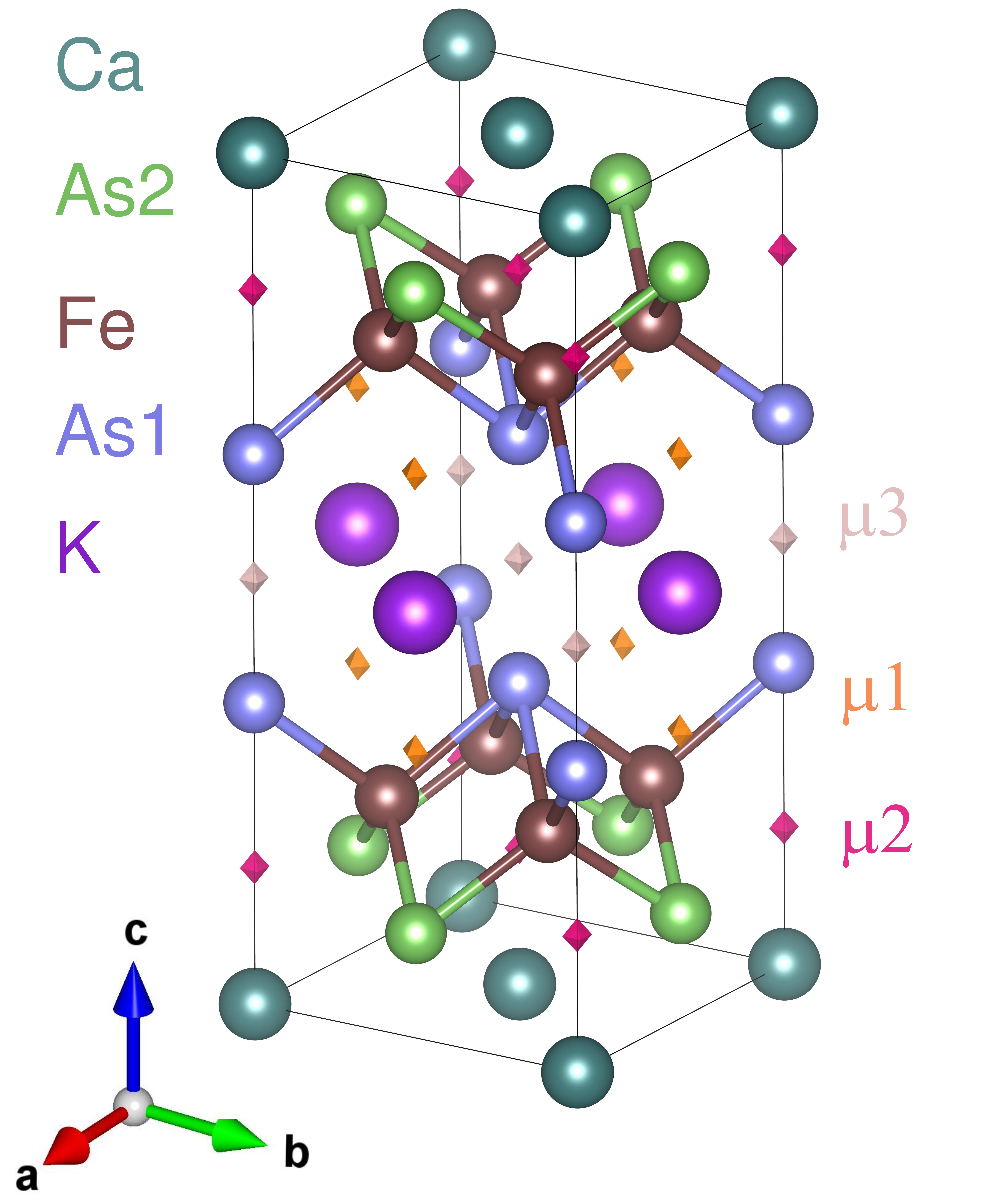}
% \vspace{-1.0cm}
%
\caption{ The crystal structure of CaKFe$_4$As$_4$ and the muon stopping sites $\mu1$, $\mu2$ and $\mu3$ within the tetragonal subgroup $P4/mbm$ (N127) of the space group $P4/mmm$ (N123). The $P4/mbm$ subgroup has the same origin as the parent group $P4/mmm$ and the basis $(a, b, c)$ rotated in the $ab-$plane on $45^{\rm o}$ compared to the basis $(a', b', c')$ of the $P4/mmm$. Atoms and muon stopping sites are at the positions: Ca -- $2a$ (0,0,0); K -- $2c$ (0,0.5,0.5); As1 -- $4e$ (0,0,0.35408), As2 -- $4f$ (0,0.5,0.12310); Fe -- $8k$ (0.25, 0.75, 0.76800); $\mu1$ -- $4f$ (0,0.5, 0.301); $\mu2$ -- $4e$ (0,0, 0.186); and $\mu3$ -- $2b$ (0,0,0.5). The crystal structure is visualized by using VESTA package, Ref.~\onlinecite{Vesta}. }
 \label{fig:crystal-structure_P4mbm}
\end{figure}

The dipolar field analysis follows the procedure described in the Supplemental Material part of Ref.~\onlinecite{Mallett_EPL_2015} and in the Ref.~\onlinecite{Sheveleva_Arxiv_2020}.

The single-{\it Q} and double–{\it Q} magnetic structures which can arise in the $P4/mmm$ setting with $\bm{Q}_1= (1/2, 1/2, 0) $ and $\bm{Q}_2=(-1/2, 1/2, 0)$ were considered.
For convenience, the crystal structure of CaKFe$_4$As$_4$ was described within the space group $P4/mbm$ (N127) [the subgroup of indexes 2 of the parent group $P4/mmm$ (N123) with doubled unit cell]. The $P4/mbm$ subgroup has the same origin as the parent group $P4/mmm$ and the basis $(a, b, c)$ rotated in the $ab-$plane on $45^{\rm o}$ compared to the basis $(a', b', c')$ of the $P4/mmm$, so that $a=b=2a'$ and $c=c'$ (see Fig.~\ref{fig:crystal-structure}). In $P4/mbm$ setting, the muon sites become: $\mu1$ -- $4f$ (0,0.5, 0.301); $\mu2$ -- $4e$ (0,0, 0.186); and  $\mu3$ -- $2b$ (0,0,0.5).  The primitive cell of CaKFe$_4$As$_4$ and descriptions of all ion and muon positions in $P4/mbm$ setting are given in Fig.~\ref{fig:crystal-structure_P4mbm}.
Note that within the $P4/mbm$ representation used here the magnetic and the crystallographic structure share the similar unit cell. In such a case the symmetry treatment for the Fe- and muon-site magnetic representations could be performed for the propagation vector $\bm{K}_0=(0,0,0)$.

\begin{table*}[htb]
\caption{\label{tab:table2} The results of dipolar field calculations for various spin-vortex-crystal (SVC) magnetic structures of CaKFe$_4$As$_4$. $P4/mbm$ crystal structure with $\bm{K}_0=(0;0;0)$ magnetic propagation vector was considered. The values of fields at the muon-sites were calculated by  means of Eqs~\ref{eq:Field_mu1}, \ref{eq:Field_mu2}, and \ref{eq:Field_mu3}, assuming that each Fe carries the magnetic moment of 1~$\mu_{\rm B}$.  }
\begin{tabular}{c|c|c|cc|cc|ccccc}
\hline
\hline
& & &\multicolumn{2}{c|}{$\mu 1$ ($4f$)}& \multicolumn{2}{c|}{$\mu 2$ ($4e$)}&\multicolumn{2}{c}{$\mu 3$ ($2b$)}\\  \cline{4-9}
IR&Fe&SVC&Field&Field Value&Field&Field Value&Field&Field Value \\
&order parameter&motif&symmetry&for $1\mu_{\rm B}$/Fe  & symmetry&for $1\mu_{\rm B}$/Fe  & symmetry&for $1\mu_{\rm B}$/Fe  \\
&&&&(mT)&&(mT)&&(mT)\\
\hline
$\tau_1-A_{\rm 1g}$&$L_{3x}^{(-)}+L_{1y}^{(-)}$&hedgehog &--&0&$L_{2z}^{(I)}$&365.58&$l_z$&94.36\\
$\tau_2-A_{\rm 1u}$&$L_{1x}^{(+)}-L_{3y}^{(+)}$&loop    &$L_{3z}^{(I)}$&0&$L_{3z}^{(I)}$&0&--&0\\
$\tau_3-A_{\rm 2g}$&$L_{1x}^{(-)}-L_{3y}^{(-)}$&loop    &$F_{3z}^{(II)}$&0&$F_{z}^{(I)}$&0&$m_z$&0\\
$\tau_4-A_{\rm 2u}$&$L_{3x}^{(+)}+L_{1y}^{(+)}$&hedgehog &--&0&$L_{1z}^{(I)}$&356.04&--&0\\
$\tau_5-B_{\rm 1g}$&$L_{3x}^{(-)}-L_{1y}^{(-)}$&hedgehog &$L_{2z}^{(II)}$&417.66&--&0&--&0\\
$\tau_6-B_{\rm 1u}$&$L_{1x}^{(+)}+L_{3y}^{(+)}$&loop    &--&0&--&0&--&0\\
$\tau_7-B_{\rm 2g}$&$L_{1x}^{(-)}+L_{3y}^{(-)}$&loop    &--&0&--&0&--&0\\
$\tau_8-B_{\rm 2u}$&$L_{3x}^{(+)}-L_{1y}^{(+)}$&hedgehog &$L_{1z}^{(II)}$&414.53&--&0&--&0\\
\hline
\hline
\end{tabular}
\end{table*}

Eight different double–{\it Q} magnetic spin-vortex-crystal (SVC) structures with orthogonal Fe moments lying in the $ab-$plane were introduced. They were presented by magnetic order parameters from one dimensional irreducible representations (IR's) from $\tau_1$ to $\tau_8$. The SVC structures were divided into two groups with so-defined "hedgehog" ($\tau_1$, $\tau_4$, $\tau_5$, and $\tau_8$) and "loop" ($\tau_2$, $\tau_3$, $\tau_6$, $\tau_7$) motif in accordance with its orthogonal arrangement of the spin pattern.\cite{Meier_NPJ_2018}
In addition to double–{\it Q}, the single-{\it Q} magnetic structures with $\tau_9$ and $\tau_{10}$ IR's could be considered (not shown, see Refs.~\onlinecite{Mallett_EPL_2015,Sheveleva_Arxiv_2020} for details).
The magnetic structures corresponding to $\tau_1 -\tau_{8}$ irreversible representations are shown in Figs.~18 and 19 of Ref.~\onlinecite{Sheveleva_Arxiv_2020}.

Table~\ref{tab:table2} summarizes the outcome of the dipolar field calculations for the magnetic order parameters of CaKFe$_4$As$_4$. The linear combinations of Fe-magnetic moments [$\vec{F}^{(+),(-)}$, $\vec{L}_1^{(+),(-)}$, $\vec{L}_2^{(+),(-)}$, and $\vec{L}_3^{(+),(-)}$] and the staggered magnetic fields at muon sites ($\vec{F}^{I,II}$, $\vec{L}_1^{I,II}$, $\vec{L}_2^{I,II}$, and $\vec{L}_3^{I,II}$) are described in Ref.~\onlinecite{Sheveleva_Arxiv_2020}.
The values of magnetic fields at muon-stopping sites were calculated by means of Eqs.~\ref{eq:Field_mu1}, \ref{eq:Field_mu2} and \ref{eq:Field_mu3}.  Values of Fe moments were set to 1~$\mu_{\rm B}$.

Fields at $\mu 1$ ($4f$) muon-sites with coordinates $(0, 0.5, 0.301)$ are:
\begin{widetext}
\begin{eqnarray}
   \begin{pmatrix}
   B_x  \\
   B_y  \\
   B_z  \\
   \end{pmatrix}
&=&
   \begin{pmatrix}
   0 & 0 & 0  \\
   0 & 0 & 295.33  \\
   0 & 295.33 & 0  \\
   \end{pmatrix}
   \begin{pmatrix}
   L_{1x}^{(-)}  \\
   L_{1y}^{(-)}  \\
   L_{1z}^{(-)}  \\
   \end{pmatrix}
+
   \begin{pmatrix}
   0 & 0 & -295.33  \\
   0 & 0 & 0  \\
   -295.33 & 0 & 0  \\
   \end{pmatrix}
   \begin{pmatrix}
   L_{3x}^{(-)}  \\
   L_{3y}^{(-)}  \\
   L_{3z}^{(-)}  \\
   \end{pmatrix}
\nonumber \\
&+&
   \begin{pmatrix}
   0 & 0 & 0  \\
   0 & 0 & 293.12  \\
   0 & 293.12 & 0  \\
   \end{pmatrix}
   \begin{pmatrix}
   L_{1x}^{(+)}  \\
   L_{1y}^{(+)}  \\
   L_{1z}^{(+)}  \\
   \end{pmatrix}
+
   \begin{pmatrix}
   0 & 0 & -293.12  \\
   0 & 0 & 0  \\
   -293.12 & 0 & 0  \\
   \end{pmatrix}
   \begin{pmatrix}
   L_{3x}^{(+)}  \\
   L_{3y}^{(+)}  \\
   L_{3z}^{(+)}  \\
   \end{pmatrix}
 \label{eq:Field_mu1}
\end{eqnarray}
The field at one of $\mu 2$ ($4e$) muon-sites with coordinates $(0, 0, 0.186)$ is:
\begin{eqnarray}
   \begin{pmatrix}
   B_x  \\
   B_y  \\
   B_z  \\
   \end{pmatrix}
&=&
   \begin{pmatrix}
   0 & 0 & 0  \\
   0 & 0 & 258.51  \\
   0 & 258.51 & 0  \\
   \end{pmatrix}
   \begin{pmatrix}
   L_{1x}^{(-)}  \\
   L_{1y}^{(-)}  \\
   L_{1z}^{(-)}  \\
   \end{pmatrix}
+
   \begin{pmatrix}
   0 & 0 & 258.51  \\
   0 & 0 & 0  \\
   258.51 & 0 & 0  \\
   \end{pmatrix}
   \begin{pmatrix}
   L_{3x}^{(-)}  \\
   L_{3y}^{(-)}  \\
   L_{3z}^{(-)}  \\
   \end{pmatrix}
\nonumber \\
&+&
   \begin{pmatrix}
   0 & 0 & 0  \\
   0 & 0 & 251.76  \\
   0 & 251.76 & 0  \\
   \end{pmatrix}
   \begin{pmatrix}
   L_{1x}^{(+)}  \\
   L_{1y}^{(+)}  \\
   L_{1z}^{(+)}  \\
   \end{pmatrix}
+
   \begin{pmatrix}
   0 & 0 & 251.76  \\
   0 & 0 & 0  \\
   251.76 & 0 & 0  \\
   \end{pmatrix}
   \begin{pmatrix}
   L_{3x}^{(+)}  \\
   L_{3y}^{(+)}  \\
   L_{3z}^{(+)}  \\
   \end{pmatrix}
 \label{eq:Field_mu2}
\end{eqnarray}

Fields at $\mu 3$ ($2b$) muon-sites with coordinates $(0, 0, 0.5)$ are:
\begin{equation}
   \begin{pmatrix}
   B_x  \\
   B_y  \\
   B_z  \\
   \end{pmatrix}
=
   \begin{pmatrix}
   0 & 0 & 0  \\
   0 & 0 & -66.73  \\
   0 & -66.73 & 0  \\
   \end{pmatrix}
   \begin{pmatrix}
   L_{1x}^{(-)}  \\
   L_{1y}^{(-)}  \\
   L_{1z}^{(-)}  \\
   \end{pmatrix}
+
   \begin{pmatrix}
   0 & 0 & -66.73  \\
   0 & 0 & 0  \\
   -66.73 & 0 & 0  \\
   \end{pmatrix}
   \begin{pmatrix}
   L_{3x}^{(-)}  \\
   L_{3y}^{(-)}  \\
   L_{3z}^{(-)}  \\
   \end{pmatrix}
 \label{eq:Field_mu3}
\end{equation}
\end{widetext}

The calculations implies that magnetic structures, with the corresponding $\tau_2$ and $\tau_3$ IR's cannot be seen at any of three ($4f$, $4e$ and $2b$) muon stopping sites. The structures corresponding to $\tau_6$ and $\tau_7$ IR's may result in non-zero fields at muon sites only for a case of deviation of Fe ions from the starting $8k$ $(0.25, 0.75, z)$ position to $8k$ $(x, x+0.5, z)$ with $x\neq 0.25$. However, the x-ray diffraction studies were not detecting any lattice distortions in CaK(Fe$_{0.949}$Ni$_{0.051})_4$As$_4$ below $T_{\rm N}\simeq50$~K.\cite{Meier_NPJ_2018} The iron position $8k$ $(0.25, 0.75, z)$  is expected, therefore, to stay unchanged in the magnetic ordered state. These arguments exclude all loop-type SVC structures from the consideration. The differentiation between 4 'hedgehog' SVC magnetic structures could be further made by comparing the ZF-$\mu$SR data with the results of calculations presented in Table~\ref{tab:table2}.

The symmetry analysis presented in Table~\ref{tab:table2} could be applied to determine the hyperfine magnetic fields induced by SVC orders at As1 ($4e$) and As2 ($4f$) ions. The respective staggered magnetic fields will have the same direction and distribution for $\mu 2$ and As1 sites, as well as for $\mu 3$ and As2 sites. The As-NMR studies reveal the presence of hyperfine fields at As1 and the absence of such a field at As2 ions in the magnetically ordered state in CaK(Fe$_{1-x}$Ni$_x)_4$As$_4$.\cite{Meier_NPJ_2018} This is consistent with the $\tau_1$ or $\tau_4$ representations of the hedgehog SVC order.

\section{Machida approach \label{sec:Machida-Approach}}

The Machida's self-consistent gap equations for the coupled magnetic ($M$) and the superconducting ($\Delta$) order parameters are given in Refs.~\onlinecite{Budko_PRB_2018} and \onlinecite{Machida_JPSJ_1981} (see also Eqs.~\ref{eq:Machida1} and \ref{eq:Machida2} in Sec.~\ref{seq:T-dependece_of_m}). Here we consider the derivation of $M$ and $\Delta$ at $T=0$.

The variables used in Machida's theory are described in Sec.~\ref{seq:T-dependece_of_m}.
The pure magnetic order parameter is assumed to satisfy the BCS relation  $M_0/T_{s0}=\pi/e^\gamma\approx 1.76$ ($M_0$ is zero-temperature energy gap in the electron spectrum in the absence of superconducting order). It is convenient for our purpose to use dimensionless variables:
\begin{eqnarray}
 t=\frac{T}{T_{s0}}, \ \ \quad  \delta=\frac{\Delta}{M_0},\ \ \quad m= \frac{M }{M_0}.
  \label{eq:Machida_dimensionless-variables}
\end{eqnarray}
Note that these $m$ and $\delta$ differ from those used in Ref.~\onlinecite{Budko_PRB_2018}. This normalization is convenient because $m(0)=1$.

\subsection{Solution at $T=0$}

Considering Eq.~\ref{eq:Machida1} at $T\to 0$, the sum could be replaced with an integral according to $2\pi T\sum_\omega  \to \int_0^{\omega_{\rm s}} d(\hbar\omega)$. After integration in first two terms and summation in the last term from $n=0$ to the maximum corresponding to $\omega_s=2\pi T N_{\rm s} $, Eq.~\ref{eq:Machida1} transforms to:
\begin{eqnarray}
   \ln\frac{T}{T_{\rm s0}}&=&\frac{M+\Delta}{2 M }  \ln  \frac{2\omega_s}{M+\Delta } +\frac{M-\Delta}{2 M }  \ln  \frac{2\omega_s}{|M-\Delta|}  \nonumber \\
   && -\ln(4e^\gamma  N_s).
   \label{eq:Kogan1_first-iteration}
\end{eqnarray}
Here, the standard treatment of
\begin{equation}
 2\pi T\sum_{\omega>0}^{\omega_s} \frac{1}{\omega}=\sum_{n=0}^{N_s}\frac{1}{n+1/2}= \psi\left(\frac{3}{2}+N_s\right)-\psi\left(\frac{1}{2}\right),\qquad \nonumber
\end{equation}
for $N_s\gg 1$ is used.

With the use of dimensionless variables of Eq.~\ref{eq:Machida_dimensionless-variables}, Eq.~\ref{eq:Kogan1_first-iteration} is further transformed to:
\begin{eqnarray}
  (m+\delta) \ln(m+\delta)+  (m-\delta) \ln|m-\delta|=0.
   \label{eq:Kogan1_second-iteration}
\end{eqnarray}
In the absence of superconductivity this reduces to $\ln m=0$, {\it i.e.} $m=1$ and $M(0)=M_0$ as it should.

Similar manipulations with Eq.~\ref{eq:Machida2} give:
\begin{eqnarray}
\ln\frac{T}{T_{\rm c0}} &=& n_1  \left[ \frac{M+\Delta}{2 \Delta }  \ln  \frac{2\omega_{\rm D}}{M+\Delta }- \frac{M-\Delta}{2 \Delta }  \ln  \frac{2\omega_{\rm D}}{|M-\Delta|} \right]\nonumber\\
&& + n_2 \ln\frac{2\omega_{\rm D}} {\Delta} -\ln(4e^\gamma  N_{\rm D}).
   \label{eq:Kogan2_first-iteration}
\end{eqnarray}
where $N_{\rm D}=\omega_{\rm D}/2\pi T$. In dimensionless variables it takes the form:
 \begin{eqnarray}
&n_1&   \frac{(m-\delta )  \ln | R(m-\delta)| - (m+\delta) \ln [R (m+\delta)]}{2\delta}  \nonumber\\
- &n_2&  \ln (R\delta)  =0.
   \label{eq:Kogan2_second-iteration}
\end{eqnarray}
Here,
 \begin{eqnarray}
  R=\frac{T_{s0}}{T_{c0}} = \frac{M_0}{\Delta_0} \nonumber
\end{eqnarray}
and $\Delta_0$ is the superconducting gap in the absence of SDW order. In particular, if $n_1=0$, we have $\ln R\delta=0$, {\it i.e.}
 \begin{eqnarray}
  R\delta=\frac{M_0}{\Delta_0}  \frac{\Delta}{M_{0}} =  \frac{\Delta}{\Delta_0}=1 \nonumber
\end{eqnarray}
and $\Delta=\Delta_0$, as it should.

A more compact form of Eq.~\ref{eq:Kogan2_second-iteration} is
 \begin{eqnarray}
\delta\ln (R\delta)  +n_1[   (m+\delta) \ln   (m+\delta)  -\delta\ln \delta]   =0.
   \label{eq:Kogan2_third-iteration}
\end{eqnarray}

\begin{figure}[htb]
\centering
\includegraphics[width=0.9\linewidth]{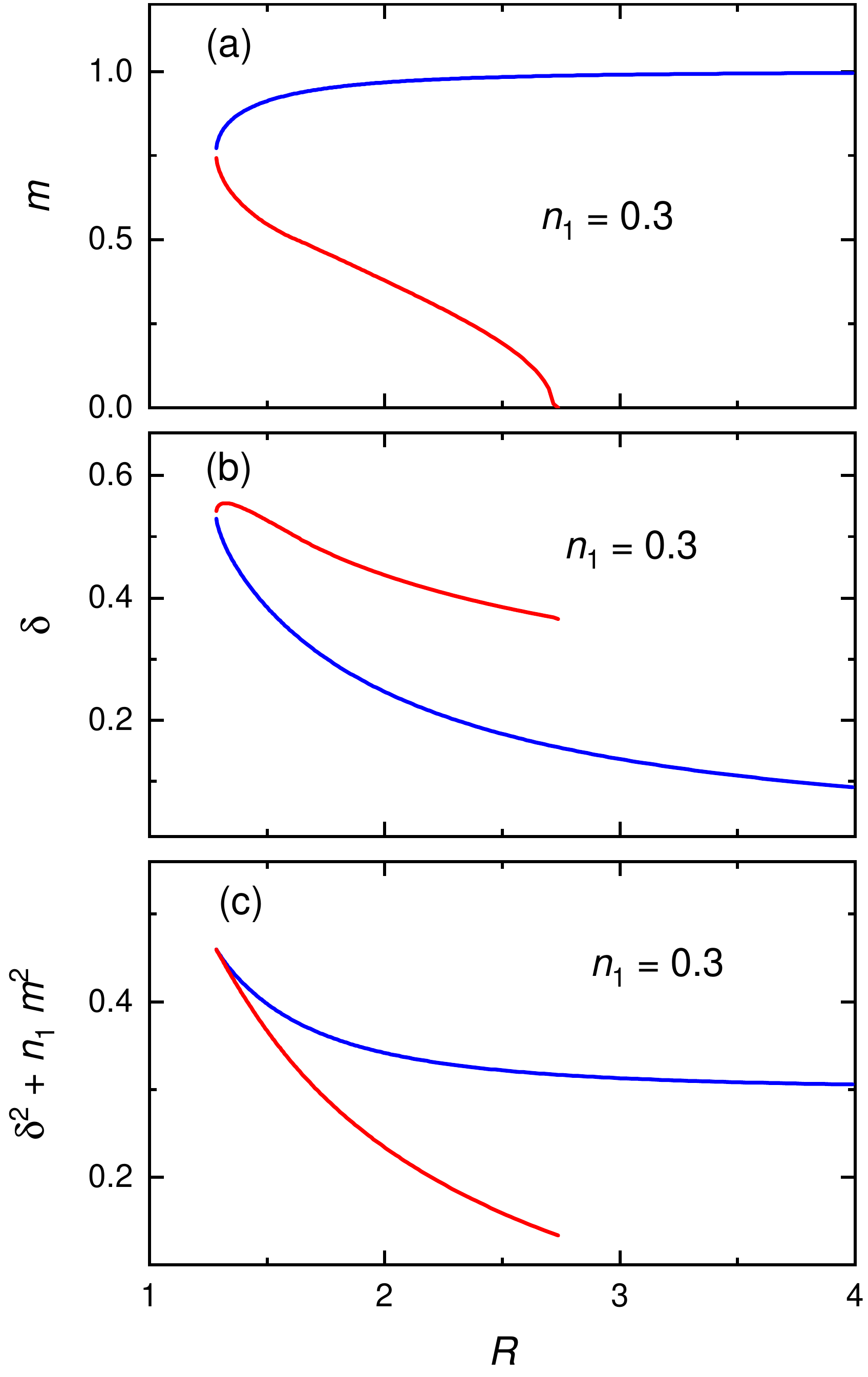}
% \vspace{-1.0cm}
%
\caption{ (a) $m$ and (b) $\delta$ as a function of the parameter $R=T_{\rm c0}/T_{\rm s0}=\Delta_0/M_0$ corresponding to solutions of the system of  Equations~\ref{eq:Kogan1_second-iteration} and \ref{eq:Kogan2_second-iteration} for $n_1=0.3$. The blue(red) branches represent solutions for $m$ increasing(decreasind) as a function of $R$.  (see text for details).  (c) The $R$ dependence of the quantity $\delta^2+n_1 m^2$ entering the condensation energy at $T=0$, Eq.~\ref{eq:Kogan_condensation-energy}. }
 \label{fig:m-delta_vs_R_n030}
\end{figure}

\begin{figure}[htb]
\centering
\includegraphics[width=0.9\linewidth]{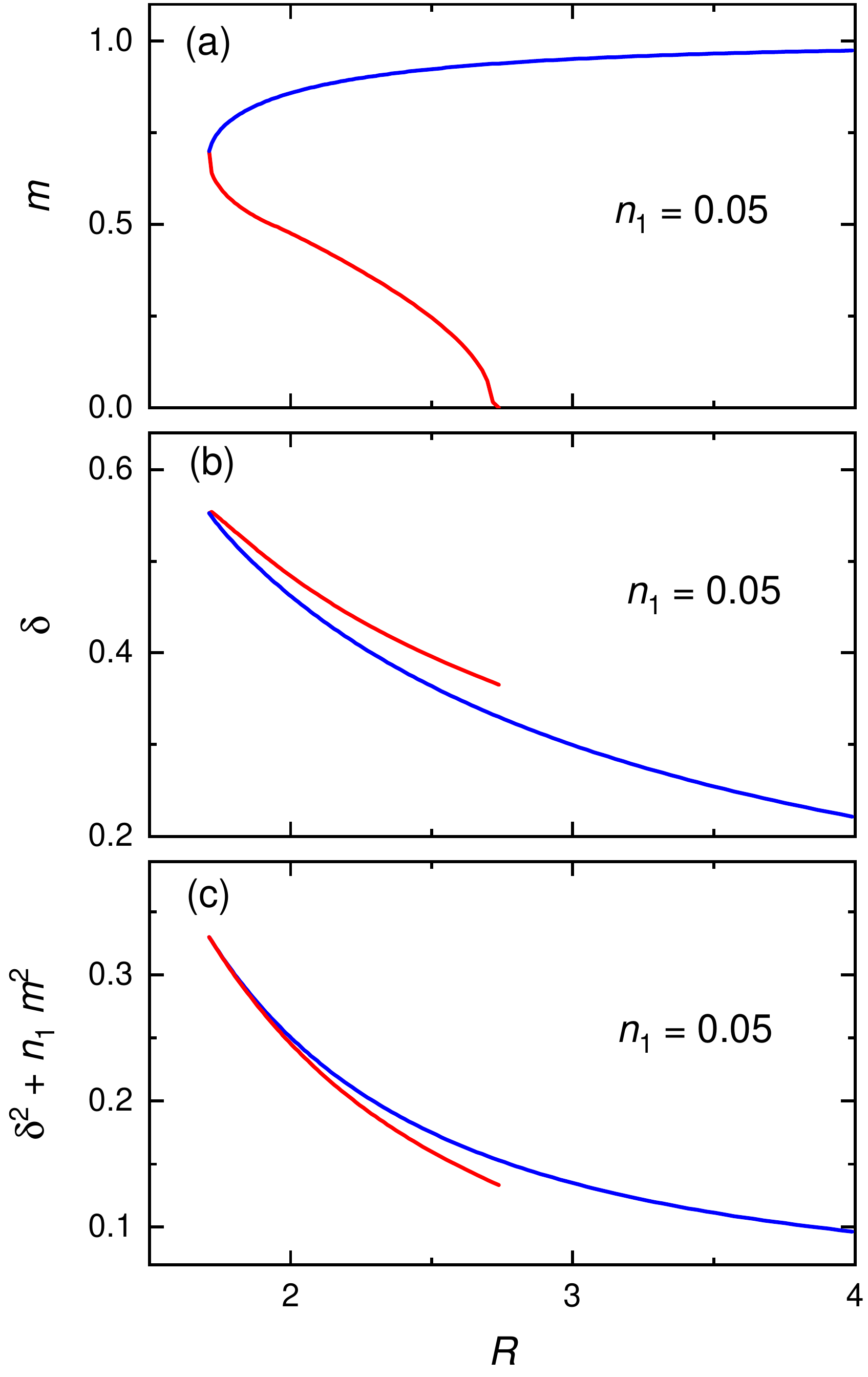}
% \vspace{-1.0cm}
%
\caption{The same as in Fig.~\ref{fig:m-delta_vs_R_n030} but for $n_1=0.05$.}
 \label{fig:m-delta_vs_R_n005}
\end{figure}

Hence, we have a system of two equations,  Eq.~\ref{eq:Kogan1_second-iteration} and Eq.~\ref{eq:Kogan2_second-iteration}, which can be solved for  $m$ and $\delta$ if $n_1$ and $R$ are given. In particular, one can fix $n_1$ and plot $m$ and $\delta$ as a function of $R$ as is done in panels (a) and (b) of Figs.~\ref{fig:m-delta_vs_R_n030} and \ref{fig:m-delta_vs_R_n005} for $n_1=0.3$ and $n_1=0.05$. Two solution branches, corresponding to the increase and decrease of $m$ with increasing $R$, are represented by blue and red color, respectively.

Similar calculations can be performed by keeping the parameter $R$ constant and evaluating $n_1$ (not shown).

\subsection{Solutions boundaries}

\begin{figure}[htb]
\centering
\includegraphics[width=0.9\linewidth]{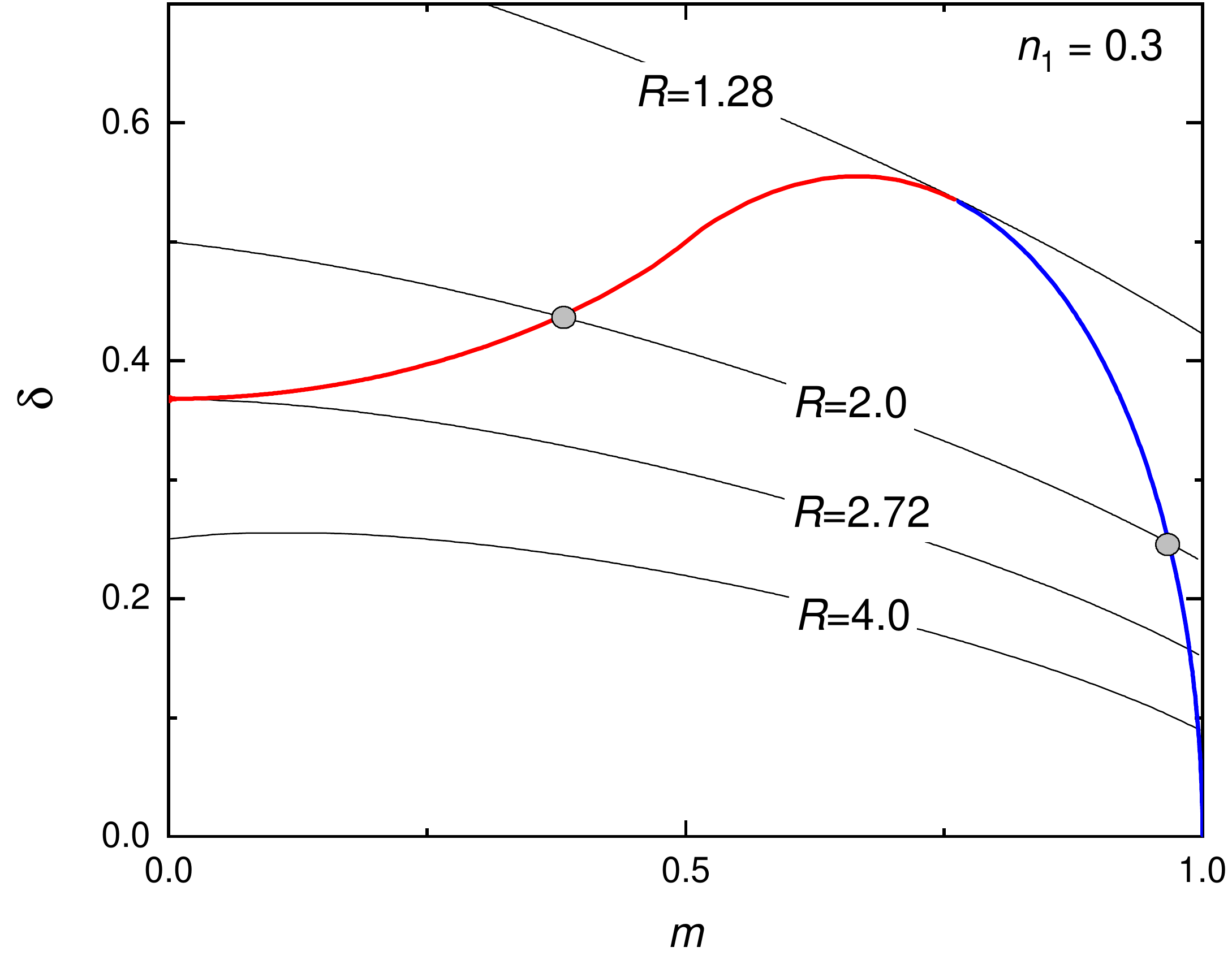}
% \vspace{-1.0cm}
%
\caption{ Solutions of Eqs.~\ref{eq:Kogan1_second-iteration} and \ref{eq:Kogan2_second-iteration}. Points ($m,\delta$) on the red/blue curve are solutions of \ref{eq:Kogan1_second-iteration}. The black curves are contours of constant $R$'s calculated with the help of Eq.~\ref{eq:Kogan2_second-iteration} (or Eq.~\ref{eq:Kogan2_third-iteration}) for $n_1=0.3$.  The points, where two curves cross, correspond to solutions of the system of equations  Eqs.~\ref{eq:Kogan1_second-iteration} and \ref{eq:Kogan2_second-iteration}.}
 \label{fig:m_vs_delta_n030}
\end{figure}

\begin{figure}[htb]
\centering
\includegraphics[width=0.9\linewidth]{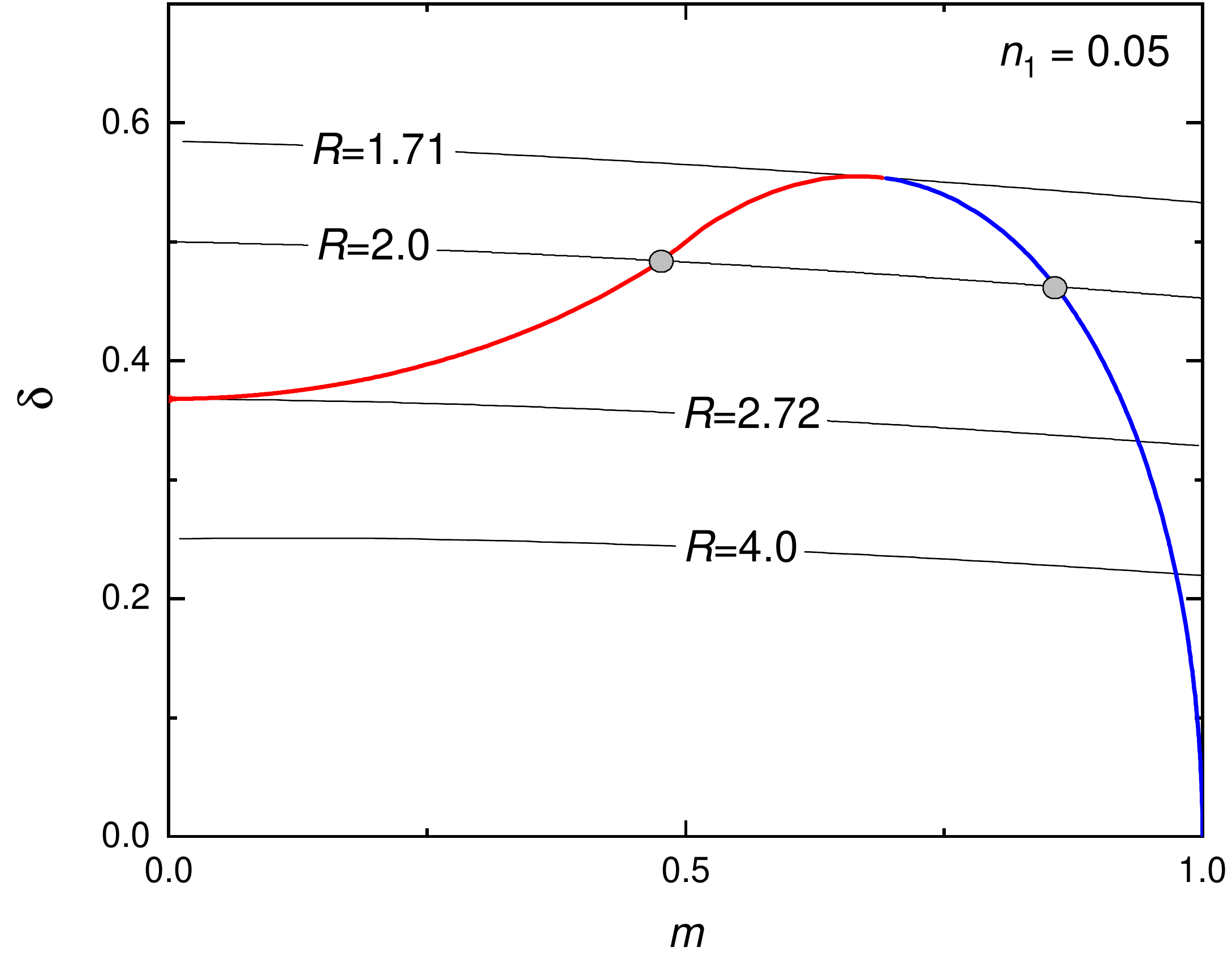}
% \vspace{-1.0cm}
%
\caption{ The same as in Fig.~\ref{fig:m_vs_delta_n030}, but for $n_1=0.05$.}
 \label{fig:m_vs_delta_n005}
\end{figure}

One can get insight to numerical procedure of solving the system of equations for $m$ and $\delta$ by noting that Eq.~\ref{eq:Kogan1_second-iteration} does not contain  material parameters $R$ and $n_1$, {\it i.e.} it is universal. Consider the left-hand site of this equation as a function  $\Phi (m,\delta)$ one may plot a contour $\Phi (m,\delta)=0$. Figs.~\ref{fig:m_vs_delta_n030} and \ref{fig:m_vs_delta_n005} represents such contours as a red/blue curve. The separation of $m(\delta)$ on the 'red' and the 'blue' branches is the same as for $m(R)$'s presented in Figs.~\ref{fig:m-delta_vs_R_n030} and \ref{fig:m-delta_vs_R_n005} for $n_1=0.3$ and 0.05, respectively.

Points $\delta$ {\it vs.} $m$ at these curve satisfy Eq.~\ref{eq:Kogan1_second-iteration}. In particular, if the superconductivity is completely suppressed, $\delta=0$, and $\Phi (m,0)=2m\ln m=0$ yields $m=1$ (the point $\delta=0$ and $m=1$   is   the right-most edge of the blue curve at Figs.~\ref{fig:m_vs_delta_n030} and \ref{fig:m_vs_delta_n005}).

In order to obtain the left edge, {\it i.e.}, the limit of small   $m$ at finite $\delta$, one expands $\Phi $ in powers of $m$:
 \begin{eqnarray}
 &(\delta+m) \left(\ln \delta+m/\delta\right) + (m-\delta) \left(\ln \delta- m/\delta\right)\nonumber\\
 &=2m(\ln \delta+1)  =0.
   \label{eq:e15}
\end{eqnarray}
Hence, at small $m$ we have $\delta=1/e\approx 0.368$, that confirms  the numerical result for the left-most edge of the red/blue curve.

The second equation of the system, Eq.~\ref{eq:Kogan2_third-iteration}, can be written as:
 \begin{eqnarray}
 R= \exp\left[-\frac{ (1-n_1) \delta\ln \delta   +n_1(m+\delta)\ln   (m+\delta)}{\delta}\right].
   \label{e15}
\end{eqnarray}
One can now plot contours of $R(m,\delta,n_1)=$ const for a fixed $n_1$, {\it e.g.} for $n_1=0.3$ (Fig.~\ref{fig:m_vs_delta_n030}), on the same graph. In such a case the solutions of both equations are at the points where the red/blue curve crosses contours of constant $R$ (see {\it e.g.} the grey points for $R=2$ curve). As expected, the contour touch the left edge of $m(\delta)$ curve corresponds to $R=e=2.718$.
Following the data presented in Fig.~\ref{fig:m-delta_vs_R_n030}, there are no solutions at all if $R \lesssim 1.28$, there is a single solution for $R\simeq 1.28$, there are two solutions for $1.28 \lesssim R<e\approx 2.72$, and only one solution if $R> e $.

Similar situation is shown in Fig.~\ref{fig:m_vs_delta_n005} for $n_1=0.05$: no solutions for $R\lesssim 1.7$, two solutions in the interval $1.7\lesssim R <e \approx 2.72$, and a single solution for $ R>e $.

  \subsection{Condensation energy}

Within Machida's model, the condensation energy of both phases at $T=0$ is
 \begin{eqnarray}
  \frac{N(0)\Delta^2(0)}{2}+ \frac{N_1M^2(0)}{2} = \frac{N(0)M_0^2 }{2}\left( \delta^2+ n_1m^2  \right).
   \label{eq:Kogan_condensation-energy}
\end{eqnarray}
Here $N_1$ is the density of states (DOS) on the Fermi surface part responsible for SDW and $N(0)$ is the full DOS.

Figs.~\ref{fig:m-delta_vs_R_n030}~(c) and  \ref{fig:m-delta_vs_R_n005}~(c) compare the condensation energies for two type of solutions as they represented by the blue and the red curves, respectively. Obviously, in the region where two solutions for a given DOS $n_1$ are possible, $(m_1,\delta_1)$ and  $(m_2,\delta_2)$, the blue curve stay above the red one. By decreasing $n_1$, the difference between them decreases and both of them will coincide at the limit of $n_1\rightarrow 0$, {\it i.e.} for a case when the magnetism completely vanishes.
The question remains whether or not the $m$ {\it vs} $\delta$ solutions corresponding to the red part of the curve are just non-physical or correspond  to something observable.


\begin{thebibliography}{99}


\bibitem{Kamihara_JAPS_2008} Y. Kamihara, T. Watanabe, M. Hirano, and H. Hosono, J. Am.Chem. Soc. {\bf 130}, 3296 (2008).

\bibitem{Hsu_PNAS_2008} F.-C. Hsu, J.-Y. Luo, K.-W. Yeh, T.-K. Chen, T.-W. Huang, P. M. Wu, Y.-C. Lee, Y.-L. Huang, Y.-Y. Chu, D.-C. Yan, and M.-K. Wu, Proc. Natl. Acad. Sci. USA {\bf 105}, 14262 (2008).

\bibitem{Stewart_RMP_2011} G. R. Stewart, Rev. Mod. Phys. {\bf 83}, 1589 (2011).

\bibitem{Chen_NSR_2014} X. Chen, P. Dai, D. Feng, T. Xiang, and F. C. Zhang, Nat. Sci. Rev. {\bf 1}, 371 (2014).

\bibitem{Paglione_NatPh_2010} J. Paglione and R. L. Greene, Nat. Phys. {\bf 6}, 645 (2010).

\bibitem{Khasanov_La1111_PRB_2011} R. Khasanov, S. Sanna, G. Prando, Z. Shermadini, M. Bendele, A. Amato, P. Carretta, R. De Renzi, J. Karpinski, S. Katrych, H. Luetkens, and N. D. Zhigadlo, Phys. Rev. B {\bf 84}, 100501(R) (2011).

\bibitem{Khasanov_FeSe_PRB_2018} R. Khasanov, R. M. Fernandes, G. Simutis, Z. Guguchia, A. Amato, H. Luetkens, E. Morenzoni, X. Dong, F. Zhou, and Z. Zhao, Phys. Rev. B 97, 224510 (2018).

\bibitem{Bednorz_ZPB_1986} J. G. Bednorz and K. A. Muller, Z. Phys. B {\bf 64}, 189 (1986).

\bibitem{Iyo_JAPS_2016} A. Iyo, K. Kawashima, T. Kinjo, T. Nishio, S. Ishida, H. Fujihisa, Y. Gotoh, K. Kihou, H. Eisaki, and Y. Yoshida, J. Am. Chem. Soc. {\bf 138}, 3410 (2016).

\bibitem{Meier_PRB_2016} W. R. Meier, T. Kong, U. S. Kaluarachchi, V. Taufour, N. H. Jo, G. Drachuck, A. E. B\"{o}hmer, S. M. Saunders, A. Sapkota, A. Kreyssig, M. A. Tanatar, R. Prozorov, A. I. Goldman, F. F. Balakirev, A. Gurevich, S. L. Bud'ko, and P. C. Canfield, Phys. Rev. B {\bf 94}, 064501 (2016).

\bibitem{Meier_PRM_2017} W. R. Meier, T. Kong, S. L. Bud'ko, and P. C. Canfield, Phys. Rev. Materials {\bf 1}, 013401 (2017).

\bibitem{Meier_PhD-Thesis_2018} William Richard Meier, PhD thesis. {\it Growth, properties and magnetism of CaKFe$_4$As$_4$}. Iowa State University, Ames, USA (2018).

\bibitem{Meier_NPJ_2018} W. R. Meier, Q.-P. Ding, A. Kreyssig, S. L. Bud’ko, A. Sapkota, K. Kothapalli, V. Borisov, R. Valenti, C. D. Batista, P. P. Orth,  R. M. Fernandes, A. I. Goldman, Y. Furukawa, A. E. B\"{o}hmer, and  P. C. Canfield,  npj Quant. Mat. {\bf 3}, 5 (2018).

\bibitem{Budko_PRB_2018} S. L. Bud'ko, V. G. Kogan, R. Prozorov, W. R. Meier, M. Xu, and P. C. Canfield, Phys. Rev. B {\bf 98}, 144520 (2018).

\bibitem{Kreyssig_PRB_2018} A. Kreyssig, J. M. Wilde, A. E. B\"{o}hmer, W. Tian, W. R. Meier, Bing Li, B. G. Ueland, Mingyu Xu, S. L. Bud'ko, P. C. Canfield, R. J. McQueeney, and A. I. Goldman, Phys. Rev. B {\bf 97}, 224521 (2018).

\bibitem{Ding_PRB_2017} Q.-P. Ding, W. R. Meier, A. E. B\"{o}hmer, S. L. Bud’ko, P. C. Canfield, and Y. Furukawa, Phys. Rev. B {\bf 96}, 220510(R) (2017).

\bibitem{Machida_JPSJ_1981} K. Machida, J. Phys. Soc. Jpn. {\bf 50}, 2195 (1981).

\bibitem{Vesta} K. Momma and F. Izumi, J. Appl. Cryst. {\bf 44}, 1272 (2011).


\bibitem{Amato_RSI_2017} A. Amato, H. Luetkens, K. Sedlak, A. Stoykov, R. Scheuermann, M. Elender, A. Raselli, and D. Graf, Review of Scientific Instruments {\bf 88}, 093301 (2017).

\bibitem{Khasanov_FeSe_int_PRB_2016} R. Khasanov, H. Zhou, A. Amato, Z. Guguchia, E. Morenzoni, X. Dong, G. Zhang, and Z. Zhao, Phys. Rev. B {\bf 93}, 224512 (2016).

\bibitem{Schenck_book_1985} A. Schenck, {\it Muon Spin Rotation Spectroscopy: Principles and Applications in Solid State Physics} (Adam Hilger Ltd., Bristol-Boston, 1985).

\bibitem{Lee_book_1999} S. L. Lee, R. Cywinski, and S. H. Kilcoyne, {\it Muon Science: Proceedings of the 51st Scottish Universities Summer School in Physics: NATO Advanced Study Institute on Muon Science}, 1728 August, 1998 (Institute of Physics, Bristol, UK, 1999).

\bibitem{Brewer_book_1994} J. H. Brewer, {\it Muon spin rotation/relaxation/resonance} in Encyclopedia of Applied Physics, edited by G. L. Trigg, Vol. 11, p. 23 (VCH, New York, 1994).

\bibitem{Yaouanc_book_2011} A. Yaouanc, and P. Dalmas de R\'{e}otier, {\it Muon Spin Rotation, Relaxation and Resonance: Applications to Condensed Matter} (Oxford University Press, Oxford, 2011).

\bibitem{MUSRFIT} A. Suter and B. M. Wojek, Phys. Procedia {\bf 30}, 69 (2012).%


\bibitem{ELK-code} "Elk code," (2009).

\bibitem{Perdew_PRB_1992} J. P. Perdew and Y. Wang, Physical Review B {\bf 45}, 13244 (1992).

\bibitem{Perdew_PRL_2008} J. P. Perdew, A. Ruzsinszky, G. I. Csonka, O. A. Vydrov, G. E. Scuseria, L. A. Constantin, X. Zhou, and K. Burke, Phys. Rev. Let. {\bf 100}, 136406 (2008).


\bibitem{Mallett_EPL_2015} B. P. P. Mallett, Y. G. Pashkevich, A. Gusev, T. Wolf, and C. Bernhard, , Europhys. Lett. {\bf 111}, 57001 (2015).

\bibitem{Sheveleva_Arxiv_2020}  E. Sheveleva, B. Xu, P. Marsik, F. Lyzwa, B. P. P. Mallett, K. Willa, C. Meingast, Th. Wolf, T. Shevtsova, Yu. G. Pashkevich, and C. Bernhard, arXiv:2004.13804.

\bibitem{Khasanov_OIE_PRL_2008} R. Khasanov, A. Shengelaya, D. Di Castro, E. Morenzoni, A. Maisuradze, I. M. Savi\'{c}, K. Conder, E. Pomjakushina, A. Bussmann-Holder, and H. Keller, Phys. Rev. Lett. {\bf 101}, 077001 (2008).


\bibitem{Khasanov_CrAs-Scirep_2015}  R. Khasanov, Z. Guguchia, I. Eremin, H. Luetkens, A. Amato, P.. K. Biswas, C. R\"{u}egg, M. A. Susner, A. S. Sefat, N. D. Zhigadlo, and  Elvezio Morenzoni  et al, Scientific Reports {\bf 5}, 13788 (2015).

\bibitem{Pratt_JPCM_2007} F. L. Pratt, P. M. Zieli\'{n}ski, M. Balanda, R. Podgajny, T.Wasiuty\'{n}ski, and B. Sieklucka, J. Phys. Condens. Matter {\bf 19}, 456208 (2007).


\bibitem{Jongh_AdPh_1974} L. J. De Jongh and A. R. Miedema, Adv. Phys. {\bf 23}, 1 (1974).


\bibitem{Klauss_JPCM_2004} H.-H. Klauss, J. Phys.: Condens. Matter {\bf 16}, S4457 (2004).

\bibitem{Maeter_PRB_2009} H. Maeter, H. Luetkens, Yu. G. Pashkevich, A. Kwadrin, R. Khasanov, A. Amato, A. A. Gusev, K. V. Lamonova, D. A. Chervinskii, R. Klingeler, C. Hess, G. Behr, B. B\"{u}chner, and H.-H. Klauss, Phys. Rev. B {\bf 80}, 094524 (2009).

\bibitem{Khasanov_MnP_PRB_2016}R. Khasanov, A. Amato, P. Bonf\`{a}, Z. Guguchia, H. Luetkens, E. Morenzoni, R. De Renzi, and N. D. Zhigadlo, Phys. Rev. B {\bf 93}, 180509(R) (2016).

\bibitem{Khasanov_MnP_JPCM_2017} R. Khasanov, A. Amato, P. Bonf\`{a}, Z. Guguchia, H. Luetkens, E. Morenzoni, R. De Renzi and N. D. Zhigadlo, J. Phys.: Condens. Matter {\bf 29}, 164003 (2017).

\bibitem{Fernandes_PRB_2010} R. M. Fernandes, D. K. Pratt,W. Tian, J. Zarestky, A. Kreyssig, S. Nandi, M. G. Kim, A. Thaler, N. Ni, P. C. Canfield, R. J. McQueeney, J. Schmalian, and A. I. Goldman, Phys. Rev. B {\bf 81}, 140501 (2010).

\bibitem{Vorontsov_PRB_2010} A. B. Vorontsov, M. G. Vavilov, and A. V. Chubukov, Phys. Rev. B {\bf 81}, 174538 (2010).

\bibitem{Schmiedt_PRB_2014} J. Schmiedt, P. M. R. Brydon, and C. Timm, Phys. Rev. B {\bf 89}, 054515 (2014).


\bibitem{Bendele_PRL_2010} M. Bendele, A. Amato, K. Conder, M. Elender, H. Keller, H.-H. Klauss, H. Luetkens, E. Pomjakushina, A. Raselli, and R. Khasanov, Phys. Rev. Lett. {\bf 104}, 087003 (2010).

\bibitem{Bendele_PRB_2010} M. Bendele, A. Ichsanow, Yu. Pashkevich, L. Keller, Th. Str\"{a}ssle, A. Gusev, E. Pomjakushina, K. Conder, R. Khasanov, and H. Keller, Phys. Rev. B {\bf 85}, 064517 (2012).

\bibitem{Goldman_PRB_2008} A. I. Goldman, D. N. Argyriou, B. Ouladdiaf, T. Chatterji, A. Kreyssig, S. Nandi, N. Ni, S. L. Bud’ko, P. C. Canfield, and R. J. McQueeney, Phys. Rev. B {\bf 78}, 100506 (2008).

\bibitem{Avci_PRB_2011} S. Avci, O. Chmaissem, E. A. Goremychkin, S. Rosenkranz, J.-P. Castellan, D. Y. Chung, I. S. Todorov, J. A. Schlueter, H. Claus, M. G. Kanatzidis, A. Daoud-Aladine, D. Khalyavin, and R. Osborn, Phys. Rev. B {\bf 83}, 172503 (2011).

\bibitem{Kim_PRB_2011} M. G. Kim, D. K. Pratt, G. E. Rustan, W. Tian, J. L. Zarestky, A. Thaler, S. L. Bud’ko, P. C. Canfield, R. J. McQueeney, A. Kreyssig, and A. I. Goldman, Phys. Rev. B {\bf 83}, 054514 (2011).

\bibitem{Kreyssig_PRB_2010} A. Kreyssig, M. G. Kim, S. Nandi, D. K. Pratt, W. Tian, J. L. Zarestky, N. Ni, A. Thaler, S. L. Bud’ko, P. C. Canfield, R. J. McQueeney, and A. I. Goldman, Phys. Rev. B {\bf 81}, 134512 (2010).

\bibitem{Wang_PRL_2012} P. Wang, Z. M. Stadnik, J. Zukrowski, A. Thaler, S. L. Bud’ko, and P. C. Canfield, Phys. Rev. B {\bf 84}, 024509 (2011).

\bibitem{Luo_PRL_2012} H. Luo, R. Zhang, M. Laver, Z. Yamani, M. Wang, X. Lu, M. Wang, Y. Chen, S. Li, S. Chang, J. W. Lynn, and P. Dai, Phys. Rev. Lett. {\bf 108}, 247002 (2012).

\bibitem{Nandi_PRL_2010} S. Nandi, M. G. Kim, A. Kreyssig, R. M. Fernandes, D. K. Pratt, A. Thaler, N. Ni, S. L. Bud’ko, P. C. Canfield, J. Schmalian, R. J. McQueeney, and A. I. Goldman, Phys. Rev. Lett. {\bf 104}, 057006 (2010).

\bibitem{Marsik_PRL_2010} P. Marsik, K. W. Kim, A. Dubroka, M. R\"{o}ssle, V. K. Malik, L. Schulz, C. N. Wang, C. Niedermayer, A. J. Drew, M. Willis, T. Wolf, and C. Bernhard, Phys. Rev. Lett. {\bf 105}, 057001 (2010).

\bibitem{Christianson_PRL_2009} A. D. Christianson, M. D. Lumsden, S. E. Nagler, G. J. MacDougall, M. A. McGuire, A. S. Sefat, R. Jin, B. C. Sales, and D. Mandrus, Phys. Rev. Lett. {\bf 103}, 087002 (2009).

\bibitem{Pratt_PRL_2011} D. K. Pratt, M. G. Kim, A. Kreyssig, Y. B. Lee, G. S. Tucker, A. Thaler, W. Tian, J. L. Zarestky, S. L. Bud’ko, P. C. Canfield, B. N. Harmon, A. I. Goldman, and R. J. McQueeney, Phys. Rev. Lett. {\bf 106}, 257001 (2011).

\bibitem{Bernhard_PRB_2012} C. Bernhard, C. N. Wang, L. Nuccio, L. Schulz, O. Zaharko, J. Larsen, C. Aristizabal, M. Willis, A. J. Drew, G. D. Varma, T. Wolf, and Ch. Niedermayer, Phys. Rev. B {\bf 86}, 184509 (2012).

\bibitem{Li_PRB_2012} Z. Li, R. Zhou, Y. Liu, D. L. Sun, J. Yang, C. T. Lin, and Guo-qing Zheng, Phys. Rev. B {\bf 86}, 180501(R) (2012).

\bibitem{Zhou_NatCom_2013} R. Zhou, Z. Li, J. Yang, D. L. Sun, C. T. Lin, and Guo-qing Zheng,  Nature Communications {\bf 4}, 2265 (2013).

\bibitem{Materne_PRB_2015} P. Materne, S. Kamusella, R. Sarkar, T. Goltz, J. Spehling, H. Maeter, L. Harnagea, S. Wurmehl, B. B\"{u}chner, H. Luetkens, C. Timm, and H.-H. Klauss, Phys. Rev. B {\bf 92}, 134511 (2015).

\end{thebibliography}
\end{document}